\documentclass[journal,peerreview,twocolumn]{IEEEtran}

\usepackage[pdftex]{graphicx}
\usepackage[dvipsnames]{xcolor}
\usepackage[noadjust]{cite}
\usepackage{amsmath}
\usepackage{algorithmic}
\usepackage{array}
\usepackage{stfloats}
\usepackage{multirow}
\usepackage{multicol}
\usepackage{slashbox,pict2e}
\usepackage{colortbl}
\usepackage{adjustbox}
\usepackage[pdftex]{lscape} 
\usepackage{longtable}
\usepackage{caption}
\usepackage{apacite}
\usepackage{balance}
\renewcommand{\thetable}{\arabic{table}}
\usepackage{setspace}
\usepackage{amssymb}
\usepackage{marvosym}

\makeatletter
\newcommand{\thickhline}{%
    \noalign {\ifnum 0=`}\fi \hrule height 1pt
    \futurelet \reserved@a \@xhline
}
\newcolumntype{"}{@{\hskip\tabcolsep\vrule width 1pt\hskip\tabcolsep}}
\makeatother

\begin{document}
\title{ 

Converging Measures and an Emergent Model: A Meta-Analysis of Human-Automation Trust Questionnaires} 

\author{\IEEEauthorblockN{Yosef S. Razin$^*$ \& Karen M. Feigh}\\
\IEEEauthorblockA{School of Aerospace Engineering\\
Georgia Institute of Technology\\
Atlanta, Georgia 30332--0250\\
$^*$Email: yrazin@gatech.edu}
}

\maketitle

\begin{minipage}{\textwidth}
\centering
\textit{Running Head:} A META-ANALYSIS OF HAI TRUST QUESTIONNAIRES\\
\textit{Manuscript Type:} Review Article\\
\textit{Acknowledgements:} Address correspondence to Yosef Razin, Georgia Institute of Technology, Atlanta, GA; yrazin@gatech.edu
\end{minipage}

\clearpage


\begin{minipage}{\textwidth}
\begin{center}
    
    \large\textbf{Abstract Page}
\end{center}

A significant challenge to measuring human-automation trust is the amount of construct proliferation, models, and questionnaires with highly variable validation.  However, all agree that trust is a crucial element of technological acceptance, continued usage, fluency, and teamwork.  Herein, we synthesize a consensus model for trust in human-automation interaction by performing a meta-analysis of validated and reliable trust survey instruments.  To accomplish this objective, this work identifies the most frequently cited and best-validated human-automation and human-robot trust questionnaires, as well as the most well-established factors, which form the dimensions and antecedents of such trust.  To reduce both confusion and construct proliferation, we provide a detailed mapping of terminology between questionnaires.  Furthermore, we perform a meta-analysis of the regression models that emerged from those experiments which used multi-factorial survey instruments.  Based on this meta-analysis, we demonstrate a convergent experimentally validated model of human-automation trust.  This convergent model establishes an integrated framework for future research.  It identifies the current boundaries of trust measurement and where further investigation is necessary.  We close by discussing choosing and designing an appropriate trust survey instrument.  By comparing, mapping, and analyzing well-constructed trust survey instruments, a consensus structure of trust in human-automation interaction is identified.  Doing so discloses a more complete basis for measuring trust that is widely applicable.   Given the increasingly recognized importance of trust, especially in human-automation interaction, this work leaves us better positioned to understand and measure it.\\

\textbf{Key Words\textemdash}
Trust in Automation, Shared Mental Models, Human-Automation Interaction, Trust Measurement and Research\\

\textbf{Pr\'ecis: }We present a new approach to achieving clarity in human-automation trust by synthesizing the factors that emerge from reliable and validated survey instruments.  A further meta-analysis of these factors reveals that an experimental consensus is finally developing around the cognitive structure of trust in technology.

\end{minipage}

\clearpage

\section{Introduction and Background}
 Though still fractured into multiple sub-fields and suffering from construct proliferation, after three decades of sustained research, a complete model is finally emerging for human-robot and human-automation trust.  The past few years have proven particularly fruitful, yielding many models, measurement instruments, and meta-reviews.  However, much research in these areas is heavily siloed, such that established researchers tend to use assessments that are familiar, and new researchers seeking a trust assessment are often at a loss, trying to make sense of the various models and distinct terminology used by each sub-field and even specific research groups.  The greatest challenge they face is which survey instrument to use to measure trust and how to assess its validity.  This gap in standards has led many to create their own instruments, making cross comparisons difficult.  
 
 In response, numerous attempts have been made to bring order to the chaos. Most notably, \citeA{McKnight2001} performed a meta-review to create a trust typology for e-commerce, which later developed into their work on trust in specific technology \cite{McKnightD2011}.  \citeA{Lee2004} produced a seminal review of trust in automation, sorting and categorizing trust definitions and their keywords, and presented an integrated model of trust calibration over time.  \citeA{Gefen2003} presented an extensive mapping of trust conceptualizations in e-commerce and then proceeded to validate an integrative model experimentally.  Further systematic reviews of trust in automation models and findings were performed by \citeA{Hoff2015} and  \citeA{Hancock2011a}, the latter including a meta-analysis of results, which was recently updated \cite{Hancock2020}.  A general review of trust measurement was presented by \citeA{french2018trust}, and a small qualitative meta-analysis of trust survey instruments has also been presented \cite{brzowski2019trust}.  Beyond these critical works were also dozens of papers and dissertations that made varying contributions  \cite<e.g.,>{schaefer2013perception,Yagoda,Desai2012,helldin2014transparency,Ross2008,Hoffman2013}.
 
 While each of these works made significant contributions, it is clear that much work remains.  Many assessments still focus on fulfilling niche requirements, and there is limited cross-communication between sub-fields.  Results are often mixed, and the way forward remains unaddressed.  Progress has been most concrete in the development of survey instruments \cite{McKnightD2011,Gefen2003,schaefer2013perception}, but empirical validation is spotty, and construct proliferation has stifled progress.
 
 This paper proposes a way forward, leveraging the reliability and validation analysis of the existing survey instruments,  to demonstrate the current convergent, established state of the overall field thus far.  We take a similar approach to that performed for human-human trust by \citeA{mcevily2011measuring} and utilize a meta-analysis.  The meta-analysis will 1) identify the most cited survey instruments and examine their validity and reliability, 2) identify the best-validated instruments, 3) identify trust-related factors within survey instruments that have shown reliability and validity and provide a comprehensive mapping of terminology between these factors, 4) establish the internal model from which trust is composed and how the various factors influence one another from the meta-analysis data, 5) provide guidance on choosing and designing future assessments of trust and, finally, 6) outline what the next steps are in validating the overall model and where we go from here.  

 Many trust papers at this point review the multitude of trust definitions \cite{schaefer2013perception,Gefen2003,mcknight2001while,muir1996trust}.   As this paper aims to identify the converging validated concept beneath human-machine trust from the current survey instruments, we leave the definition as open as possible.  Trust is a \textit{state effectuated by the trustor in which the trustee has power over some subset of the trustor's goals that the trustor believes they could not accomplish with a better net outcome on their own}\label{def:trust}.  The \textit{state} may be an attitude \cite{Merritt2008,Lee2004}, belief \cite{Gefen2003}, expectation \cite{muir1996trust}, judgement \cite{Merritt2008,Lee2004}, willingness \cite{Madsen2000,SATI,mcknight2001while,Korber2018}, confidence \cite{Madsen2000,SATI}, or reliance \cite{Hancock2011}. The perceived \textit{power} of the trustee over the trustor's goals is often expressed in terms of the trustor's vulnerability, uncertainty, or risk \cite{schaefer2013perception}.

Here we take a bottom-up approach to demonstrate what factors and interactions influence
trust and its correlates and how to measure them.  Only then will we attempt to extract what parts are trusted directly instead of antecedents or correlates.  When needed for clarity, we will note when factor names differ from standard definitions or are used in non-standard ways.
 
Thus, this work aims to cover the cognitive structure underlying trust.  By cognitive, we do not mean capability-based trust but trust as it is explicitly and consciously conceived of by participants and, therefore, amenable to self-report.  Thus, for example, we are not looking at the appearance of the robot per se but the human's perception of the robot, e.g., that it appears friendly or that they are comfortable around it.  Design, appearance, and other external factors certainly shape trust, but we focus only on identifiable internal cognitive factors that compose trust.  This focus on \emph{internal}, cognitive trust establishes a stronger foundation for understanding how and why external factors affect trust.

Trust is not just about a set of beliefs, an attitude, or an expectation but also about their calibration \cite{Lee2004}.  Additionally, these continuous epistemic beliefs or expectations may result in discrete decisions using thresholds or constraints \cite{Lee_94,Razin2021b}.  Calibration will be addressed tangentially in our discussion of Intention to Trust and External Learned Trust beliefs toward the end.  The mechanism by which these continuous beliefs collapse into discrete decisions is out of this work's scope, but a possible way to close that gap is also left to our final discussion.

This work begins by reviewing survey instruments that explicitly capture trust purposefully sourced from various related fields.  Classically, the most basic way to assess trust was to ask whether something or someone was trusted (binary/discrete) or how much it was to be trusted (continuous) \cite{Abbass,Cho2015,Ajenaghughrure2020,Cohen1998,Lewicki1998}.  As we will show, the one-item approach is still quite common.  Following the works of \citeA{Lee_94} and \citeA{Lewicki1998}, came the rise of two-item trust scales.  The former was concerned with self-confidence vs. trust in automation \cite{Lee_94}, while the latter dissociated trust and distrust and called for their independent measurement \cite{Lewicki1998}.  Both approaches remain popular, requiring two Likert scale questions and having been well-validated and replicated many times.  While at least \citeA{Lewicki1998} hints at trust being at least bi-dimensional, neither scale sheds much light on what composes an internal psychological model of trust as much as it defines what it is not.

Beyond one- and two-item assessments, alternatives exist for classifying more complex trust survey instruments.  Some try to create a single scale to just capture (dis)trust as opposed to those that posit multiple layers (vertical) \cite{merritt2019automation,merritt2011affective,McKnightD2011,SATI} or constructs (horizontal) \cite{muir1996trust,Madsen2000,Lee_94,Lee2015}.  Among multi-dimensional survey creators, some are focused on performance-based trust \cite<e.g.,>{Madsen2000,Wojton2020}, and others are relation-based (sometimes termed \textit{affective}) \cite<e.g.,>{Wechsung2013,Rupp2016}, with a recent trend towards more mixed approaches \cite<e.g.,>{McKnightD2011,Gefen2003,Wang2005,Park2020}, as described in \citeA{law2021trust}.  Finally, some aim to assess dispositional trust, as defined by \citeA{Hoff2015}, looking at factors such as faith in people or general fear of robots or technology.  On the other hand, others try to assess trust in a specific person, technology, or team.  Considering these methodological and conceptual differences as we review the survey instruments is worthwhile.

As trust measurement expanded from discrete to continuous and from one-item to multi-factor, a pattern is emerging.  While researchers may deliberately choose how to define and construct trust and capture specific aspects of it depending on the application, we hypothesize that they all are facets of a common internal psychological model.  This paper, therefore, focuses on capturing this latent multi-factorial, multi-dimensional \emph{internal} trust.  That being said, measuring just one or two aspects of trust or using fewer questions can be appropriate, depending on the context of the interaction.  Therefore, we conclude this paper by discussing how to choose an appropriate survey instrument or measure of trust.



\section{Methodology for Assessing Instrument Quality}
In order to determine instrument quality, we wanted a method that was flexible enough to accommodate various aspects of trust, multiple styles of instrument and assessment, but still retained reasonable and acceptable standards in the field of psychometric survey creation.   To this end, every survey identified was assessed for various forms of \textit{reliability} and \textit{validity} and their \textit{construction quality}.

\subsection{Reliability}\label{sec:rel}
Reliability is the consistency of a test, such as between experimenters (inter-rater), between points in time (test-retest), between versions of the same test (parallel forms), and between items that purport to measure the same thing (internal consistency). 
The primary type of reliability we are concerned with here is \textit{internal consistency}; when multiple questions are asked about the same construct, the responses should be consistent with each other.  That being said, if an assessment reported other forms of reliability, it was given credit for having a reliability assessment as other forms of reliability ideally should also be tested and reported.

Within internal consistency, several measures are available, each with its own assumptions that must be acknowledged.  The various methods of establishing internal consistency are based on different ways of measuring the correlations between item scores.  The simplest is the average inter-item correlation, which should generally be between 0.15-0.50.  Correlations less than this indicate that the measure is not homogenous (unidimensional);  correlations with more than this and the measures may have too much redundancy.  A second method is by using split-half reliabilities.  That is, the questions can be divided into two groups, and participants' scores in one group of questions can be correlated with the scores among the other group.  Of course, there are many ways to divide the questions randomly, so one might use the minimum, average, or maximum across all possible splits.  The mean of all split-half reliabilities is better known as Cronbach's $\alpha$.  Additionally, Guttman's $\lambda$-6 (G6) and McDonald's $\omega$ are also available methods to measure internal consistency.

\subsubsection{Cronbach's $\alpha$}
 While Cronbach's $\alpha$ is the standard for reporting internal consistency, it makes a considerable assumption concerning $\tau$-equivalence.  In other words, it assumes that factor loadings are equivalent across all survey items, which is not the case for most of the reported scales \cite{Flora2020}.  Even more critical for avoiding bias in $\alpha$ is that the scale is homogeneous (contains a single factor)\cite{Flora2020}, that the errors for each item are uncorrelated \cite{Flora2020}, and that there are no missing data, outliers, heavy skew, or other substantial deviations from non-normality \cite{Sheng2012,Zhang2016,Flora2020}.  If missing data or non-normality, especially outliers, are concerns, alternative robust calculations for $\alpha$ should be used \cite{Zhang2016}.  Care must also be taken to standardize $\alpha$ if different items are measured on different scales.  Even if all of these assumptions are met, $\alpha$ becomes inflated as the number of items increases.

Taking all this into account, if Cronbach's $\alpha$ is still used, it should \emph{at least} be $>0.7$ for an experimental scale or $>0.8$ for a well-established scale \cite{NajeraCatalan2019}.  It should also remain less than $<0.95$, or even $<0.9$ \cite{tavakol2011making} as too high an alpha indicates redundancy, which can lead to problems with variance inflation if a factor analysis is to be performed.

\subsubsection{Guttman's $\lambda$-6 (G6)}
While Cronbach's $\alpha$ can be understood as the ratio of individual to total variance, Guttman's $\lambda$-6 (G6) compares the total variance to the squared variance of the errors.  It presents a lower bound on the communality of each item.  While including more items generally improves G6, like with $\alpha$, they also positively bias the metric.  Furthermore, G6 is just as sensitive as $\alpha$ when it comes to non-homogeneity.

That being said, $\alpha$ helps assess whether a factor is unidimensional or multi-dimensional when compared to G6 and McDonald's coefficient $\omega$ (see below).  Both will be greater than $\alpha$ in the case of the former and less than $\alpha$ in the case of the latter \cite{RPsych,NajeraCatalan2019}, thus allowing us to check one of the assumptions of both $\alpha$ and G6 and an essential parameter in specifying $\omega$. 

\subsubsection{McDonald's $\omega$}
McDonald's $\omega$ is yet a 3rd way to measure internal consistency.  Compared to $\alpha$ or G6, McDonald's coefficient omega ($\omega$) is a better measure, placing a lower bound on total test reliability while making fewer assumptions.  Unfortunately, $\omega$ is under-utilized in the field of measurement in general, especially in trust measures.  Furthermore, the metric can be confusing because there are at least four types of $\omega$s and many different naming conventions.

\begin{itemize}
    \item $\omega_{t}$ or $\omega_{u}$: unidimensional, congeneric, continuous, sometimes referred to as total
    \item $\omega_{cat}$ or $\omega_{u-cat}$: unidimensional, congeneric, categorical
    \item $\omega_h$: hierarchical, multi-factorial, continuous
    \item $\omega_{gr}$: average $\omega$ across all factors
\end{itemize}

While $\omega_t$ does give a better estimate of reliability than either $\alpha$ or G6, the true power of $\omega$ is that it can better capture the reliability of factors that are better captured by using hierarchical factors.  While $\omega_t$ is the estimate of total test reliability, $\omega_h$ is the estimated variance accounted for by the general (highest-order) factor (similar to Cronbach's $\alpha$).  Since for a unidimensional factor, Cronbach's $\alpha$ will always be lower than  $\omega_{t}$, maintaining a high standard for $\alpha$ can help ensure a sufficiently high $\omega_{t}$ \cite{NajeraCatalan2019}.  $\omega_{t}$ can indicate how much error we may have in an estimate.  For instance, in binary classification at $\omega_{t}\simeq0.7$, classification error is approximately 10\%, whereas at $\omega_{t}>0.8$ error remains below 6\% \cite{NajeraCatalan2019}. 

For multi-dimensional scales, using $\omega$ is even more critical.  In that case, the recommended minimum values for total omega and the average omega overall factors are $\omega_t>0.8$ and $\omega_{gr}>0.6$, respectively.  This bound on $\omega_{gr}$ is especially important in the case of strong multi-dimensionality.  Strong multi-dimensionality is when items do not individually load highly on the general latent variable.  In contrast, in weak multi-dimensionality, each item also loads strongly on a single general latent variable.  Acceptable thresholds for $\omega_h$ can also change depending on the strength of multi-dimensionality, $\omega_h>0.65$ with stronger multi-dimensionality and $\omega_h>0.7$ when its weaker \cite{NajeraCatalan2019}.

While generally, both $\alpha$ and $\omega$ are used even with non-normal distributions, it has been found that both are not robust to samples with skew or positive kurtosis \cite{Sheng2012}.  Furthermore, outliers and leverage can play havoc with results \cite{Zhang2016}.  While more test items can somewhat improve these results, it is a greater sample size that offers more improvement \cite{Zhang2016}.  Additionally, algorithmic solutions offering more robust $\alpha$ and $\omega$ coefficients are now available to help mitigate these limitations \cite{ZhangA2016}.

We utilize these measures and the appropriate ranges to assess instrument reliability in Section \ref{sec:goodness} on page~\pageref{sec:goodness}.

\subsection{Statistical Validation}
When considering the validity of the assessment, three types of vailidity are typically reviewed: Construct Validity, Instrument Validity, and Content Validity.

The \textit{validity}, or to be more exact, \textit{construct validity}, is the degree to which ``the measure of a construct sufficiently measures the intended construct" \cite{o1998empirical}.  However, achieving that in practice is a multifaceted challenge.  First, we need to establish which construct is intended, whether it is well constructed, reflective or formative, and homogeneous or heterogeneous.  Once the construct is clarified and refined, showing that the measure captures the construct is even more complex.  It requires understanding how to select appropriate items, their wording, and how they co-vary.  Additionally, it must be established that the items accurately correlate well with the latent construct, that appropriate sample size and power were used, and that surface, content, statistical conclusion, internal, convergent, and discriminant validity. 

By measuring validity statistically, we are attempting to measure some parameter of a sample to come to a statistically sound conclusion about the population of interest as a whole.  Often we are trying to quantify the size of an effect in how a parameter differs between two groups.  Effect sizes may be of varying magnitudes and may also be more difficult to detect depending on the variability of the population.  The statistical power of a test describes the probability that it will detect an effect if it exists; hence it is the inverse of the false negative rate.  Smaller effects require higher powers to detect.  The significance level is the likelihood that a researcher will reject the population-level null hypothesis, even if true (false positive rate).  Thus, it is a measure of the strength of the evidence within the sample required to conclude that the effect is statistically significant.  Ultimately, the \textit{appropriate sample size} must be determined by the significance level and desired \textit{power}, the \textit{effect size}, and the standard deviation of the parameter within the population.  Thus, researchers need to establish the statistical level of their test and whether they have the appropriate power and sample size for the effect sizes and variability.  

These are just the initial steps to establish statistical conclusion validity, as the researcher must also take care that their interpretation is correct and that the chosen test's specific assumptions are met, such as the errors of the variables being uncorrelated.  Statistical conclusion validity, however, is insufficient to validate a survey fully, as it can leave potential mediators unaccounted for and relevant variables untested and even conceptually ungrounded.  Both instrument and internal validation are also needed.

\subsection{Instrument Validation}
In the first stages of instrument validation, we are mainly concerned with establishing and operationalizing the concept into a viable construct.  It is vital to get a clear definition of the construct, refine it against the established literature and have it reviewed by subject matter experts.  Within this definition process, the goal is to ensure that the construct is neither too broad nor narrow, positively and clearly defined and that the construct's nature, domain, entities, and dimensionality are assessed.  Not only is it crucial to understand if the construct is unidimensional (homogenous) or multi-dimensional (heterogenous) but also whether the construct is reflective or formative.  A reflective construct means that the latent construct is reflected in the items within the instrument.  Alternatively, a formative construct is defined by the items composing it.  However, it is essential to note that constructs are not inherently formative or reflective, and depend on how the construct and indicators are linked, the way the researcher understands them, and one's underlying ontological understanding of the world \cite{MacKenzie2011}.  

Beyond dimensionality and clear language, instrument designers must also carefully consider the item scale (ordinal, nominal, ratio) and types (discrete, continuous).  These will affect not only statistical analysis but also the types of questions that can be asked and the resolution at which they can be answered. 

\subsubsection{Content Validity} \label{sec:val}
In content validation, the primary goal is to ensure coverage.  The items on the survey must be drawn from a universal pool so that the entire construct is covered.  Additionally, each item should represent a facet of the content domain.  Given the difficulty of ensuring sufficient coverage from a universal pool, several techniques have been proposed.  First, this highlights the importance of a clear and positively defined construct.  From there, the initial question pool can be extended using items from previous surveys and top-down conceptual work from the literature review.  Subject matter experts (SMEs) can be leveraged, as in \cite{schaefer2013perception}, to review the quality of coverage and make sure all aspects of the intended construct are included and no others.  More quantitative techniques, such as that of Yao \cite{MacKenzie2011}, may be employed to ensure that individual items are sufficiently unique to maximize coverage while ensuring all construct dimensions are adequately covered.

\subsubsection{Construct Validity}
Returning now to construct validity, the goal is to ensure the construct is being measured and the results do not arise from methodological quirks.  While difficult to assess, many techniques can be used before and after the survey is administered to help mitigate concerns and verify techniques.


In addition to clear, unambiguous language, other steps can be taken to remove systemic bias resulting from how questions are presented.  For instance, many surveys will use a mixture of positive and negative items in the same factor to reduce acquiescence bias, if not extreme response bias \cite{schriesheim1981controlling}, both of which artificially inflate reliability metrics such as $\alpha$.  However, the use of negatively worded items introduces its own threat to response validity and construct homogeneity \cite{schriesheim1981controlling}, lowering $\alpha$ by implying multi-dimensionality exists, which is essentially an artifact of the opposed item valences.  This may be illustrative of one of the many trade-offs between types of validity that exist in survey design.  However, this problem can be mitigated to some extent by inverting the direction of the response,
which maintains the protection against acquiescence bias while seemingly serving as less of a threat to response validity \cite{barnette2000effects,merritt2015measuring}.  Factors solely established based on clustering negative vs. positive items should be handled with care.

Factors that share similar wording can present a similar challenge.  They may suffer both from acquiescence bias and the tendency of similarly worded items to cluster together more strongly, have correlated error, and thus yield an inflated $\alpha$ as well as invalid factors \cite{merritt2015measuring}.  Again, care should be taken when assessing such surveys' reported reliability and validity.

Other techniques to validate one particular method are measuring the construct using other modalities and instruments.  Together these can be analyzed through various means, such as multitrait-multimethod (MTMM) techniques and confirmatory factor analysis to verify whether convergent validity is achieved.  Conversely, constructs proposed to differ fundamentally can be tested for discriminant validity.  To better establish the construct experimentally, one can vary conditions or sample populations with known attributes to see that the construct changes appropriately \cite{Straub1989, MacKenzie2011}. 

Validity is a complex topic with many facets and requires care to design, develop, and test for.  This cannot be achieved in a single shot but requires iterative refinement of definitions, dimensions, language, items, survey structures,  statistical tests, and analysis choices, using both top-down conceptual approaches and bottom-up empirical methods.

\section{Surveying the Field: Identification of Trust Instruments}\label{sec: survey}
Now that we have established how we will assess the various trust instruments in the literature, we must now systematically identify which trust instruments are available to researchers.  In this section, we will describe a systematic method for trust assessment identification and the refinement of this list to a tractable number which are sufficiently useful/popular and well documented to perform our in-depth assessment.  

\subsection{Sampling Methodology}\label{sec:sampling}
To build a sufficient sample, we first collected all trust-related works mentioned among the latest and largest literature reviews in HCI, HRI, and HAI \cite{french2018trust, Hoffman2013,hancock2021evolving,brzowski2019trust}, resulting in 436 records.  We added to these 54 other sources that we gathered through our own literature review, especially of more recent works that the reviews may have missed.  We then continued to snowball our sample by reviewing these works, especially their background and 
method sections, and noting all potentially relevant trust studies in these fields.  Each time a paper in either literature review mentioned another method/survey instrument, whether used directly or not, that paper was added to the sample, resulting in 687 total records.  After accounting for duplication, we also excluded works where no experiment was performed (e.g., review works, theory papers), works that did not use a self-reported trust survey instrument (e.g., those that only measured behavior or physiology), and works to which we did not have full-text access.  Other criteria for inclusion were that the work was published in a peer-reviewed venue and accessible in English.  We ended up with a sample of 173 experimental works through June 2021.

\begin{figure}
    \centering
    \includegraphics[width=6in]{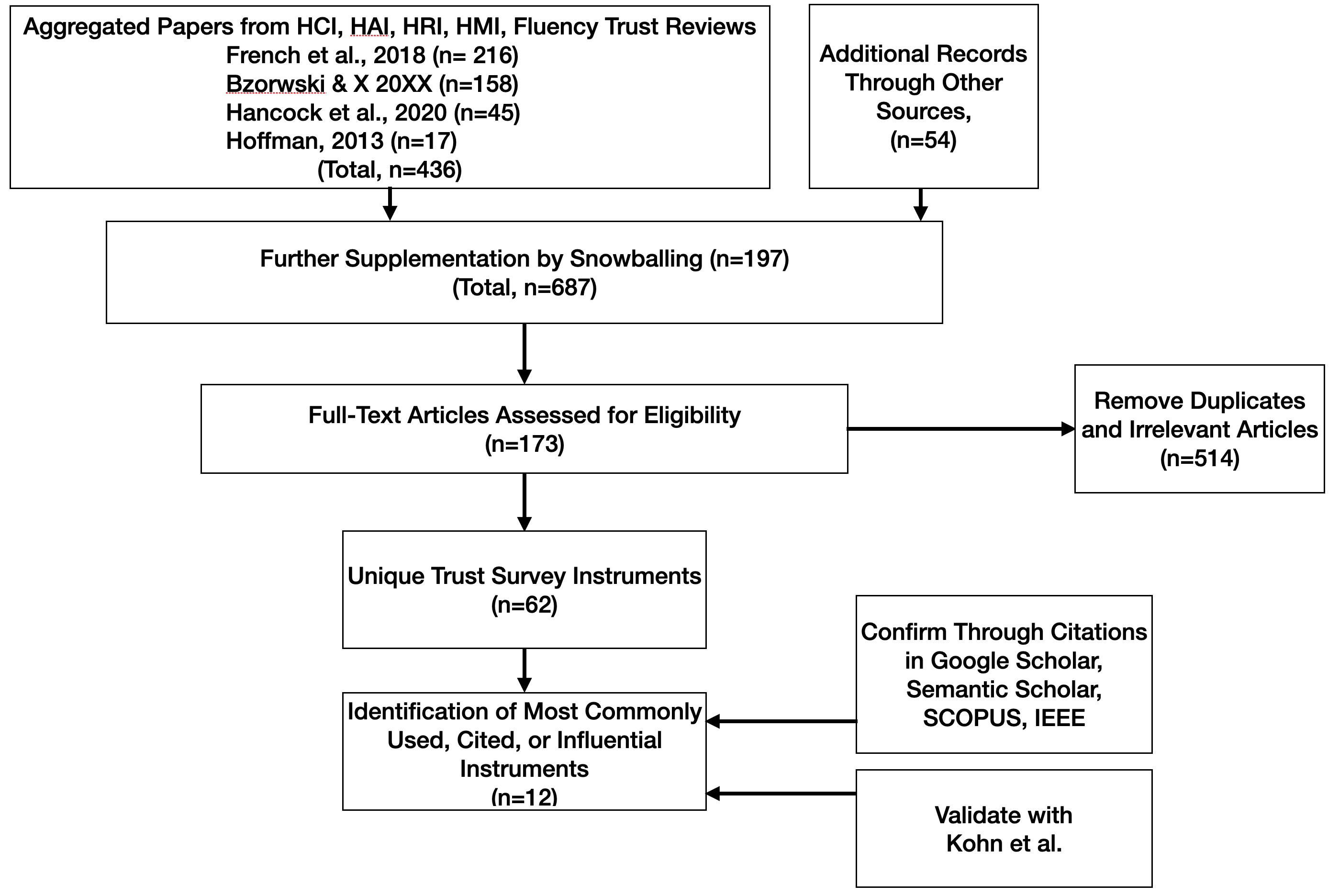}
    \caption{PRISMA diagram of literature review for identifying }
    \label{fig:3_0Prisma}
\end{figure}

Then we worked on identifying the trust instruments themselves.  Here inclusion criteria were further tightened to instruments developed and used in human-subject studies, and the whole instrument had to be accessible.  For instruments published or validated over multiple papers, relevant citations were combined across papers with duplicates removed.  In the end, we identified 62 unique survey instruments.  This process is outlined in Fig. \ref{fig:3_0Prisma}.

While we may have missed some trust instruments using this approach, it is unlikely that there remain undiscovered instruments that are widely used.  What is more likely is that instruments that are newer or less utilized, such as those primarily in use in a single lab or research group,  may have been overlooked.  However, the results produced here can be taken as indicative of the trends in the field.

\subsection{Findings and Down Selection for Analysis}\label{sec:quality}
\begin{figure}
\centering
\includegraphics[trim = 0 0 0 0, width=0.7\columnwidth]{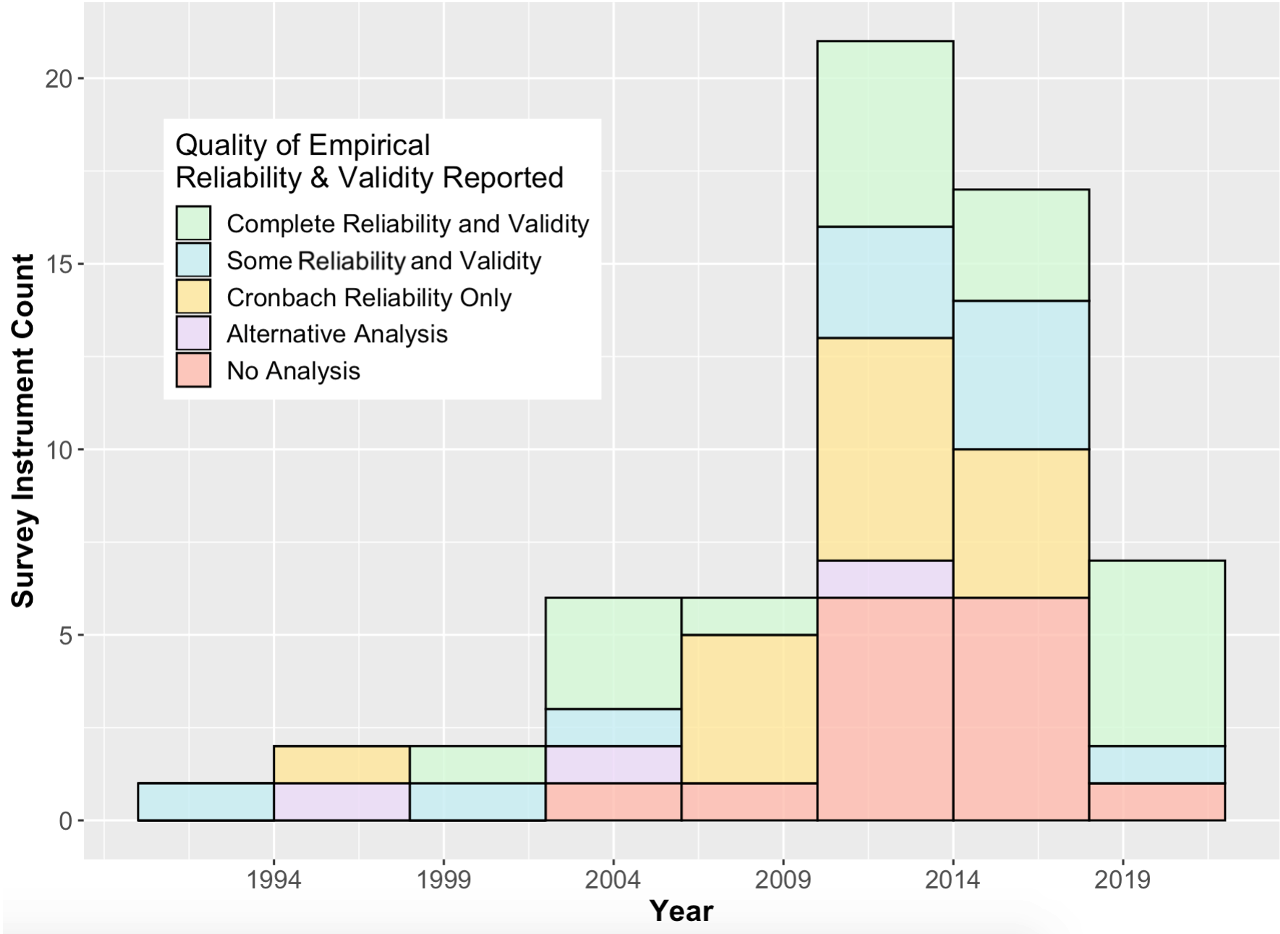}
    \caption{Number of unique multi-item ($>2$) HAI/HRI/HCI trust survey instruments by year published, broken down by the quality of empirical reliability and validity reported}
    \label{fig:hist1}
\end{figure}

\begin{figure}
    \centering
    \includegraphics[width=0.6\columnwidth]{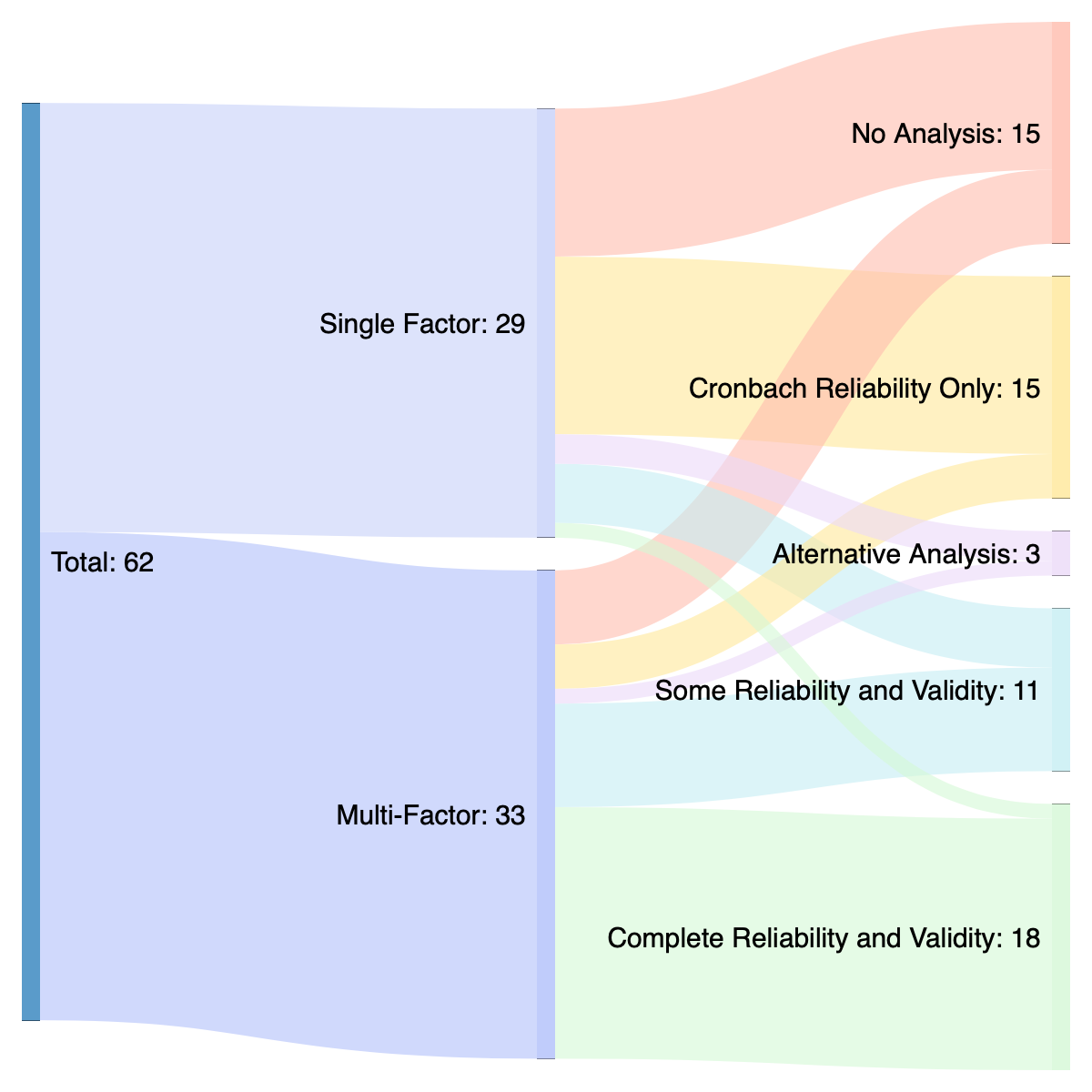}
    \caption{The categorization of HAI/HRI/HCI trust survey instruments by the quality of validation and reliability. 
    Instruments were first divided by whether they treated trust as a single factor or composed of multiple factors.  Full analysis means that a proper reliability technique was used as well as either a CFA or EFA - with their loadings, error, and fit statistics reported.}
    \label{fig:alluvial}
\end{figure}

As shown in Fig. \ref{fig:alluvial}, HMI trust instruments can be divided into two major classes, those that treat trust as a single factor and those that decompose it along multiple factors.  These classes have about equal representation within the sample.  It can also be seen that while HMI trust measurement slowly gained steam starting from the late 1980s, the past decade has seen an explosion in new trust survey instruments, as illustrated in Fig. \ref{fig:hist1}.

As illustrated in Table \ref{tab:Maj_Stats}, three very intriguing patterns emerged.  First, the single- and two-item surveys for assessing trust are very popular (12.6\% and 12.0\%, respectively), which is unsurprising given their simplicity and ease of administration and that in many fields, trust is only one of several measures being assessed in a study.  The benefits and pitfalls of such short assessments of trust are given a complete treatment by K\"{o}rber \cite{Korber2018}.  They provide a quick way to probe trust in the moment, making them suitable for inter-task assessments and tasks where time or access to training is limited.   K\"{o}rber further points out that uni-dimensional trust instruments' biggest weakness is their high variance in repeated measures, undermining their reliability.  Furthermore, they cannot explain the factors that underlie or lead to trust or lack of it.


Second, the provenance of the different instruments varied substantially.  Nearly half of the sampled experiments used one of just 12 instruments.  Fully one-third of experiments opted to develop their own survey, though often incorporating individual questions and even entire factors from the other 62.  Over one quarter used one of the other 50 identified instruments wholesale.  Thus, while the field exhibits a high degree of variability, certain instruments have proven more impactful than others. 

Third, severe divisions and terminological mismatches exist between sub-fields within trust research that have contributed to the fractiousness and confusion \cite{lewis1985trust,Gefen2003, french2018trust,krausman2022trust}.  Taken all together, these patterns indicate there is much competition and critical differences in the field, such that newcomers seeking a trust instrument may need help to choose from among the options.  This chapter will address each of these patterns in turn.    

\begin{table}[b]
    \centering
    \normalsize
    \caption{Overall statistics of the surveys used by papers included in the literature review ($n=173$). Because use could be in whole or in part, the numbers do not add up to 100\%\\  * Within which 57\% used (Lee \& Morray, 1994)}
    \begin{tabular}{lc}
      \begin{tabular}{@{}l@{}} \textbf{Instrument}\\ \textbf{Category}\end{tabular}   & \begin{tabular}{@{}c@{}}\textbf{Frequency in}\\ \textbf{Lit. Reviews}\end{tabular}  \\\hline
        Single Question & 12.6\% \\
        Two Question &  12.0\%$^*$\\
        Developed Own & 33.7\% \\
        Used Top 12 & 47.9\% \\
        Used Others & 26.9\% \\
    \end{tabular}
    \label{tab:Maj_Stats}
\end{table}

\paragraph{Most Popular Instruments.}
Twelve of the 62 instruments accounted for nearly half (47.9\%) of citations in the identified literature reviews (Table \ref{tab:Maj_Stats}), similar to the findings of Brzowski and Nathan-Roberts \cite{brzowski2019trust}.   The popularity of the majority of these instruments (8) within HRI, in particular, has been confirmed in an independent literature review \cite{kohn2021measurement}.  The pattern of citations in experimentally-focused HMI papers closely reflected the frequency of usage of each instrument in the literature review, which helps validate that our literature sample statistically reflects the experimental literature overall Table \ref{tab:overall_stats}.  Of these most used 12, McKnight's and Gefen's instruments \cite{McKnightD2011,Gefen2003} have had the most influence, with the greatest number of general citations both in background and method sections.  However, their work also goes far beyond the fields covered in our scope, covering studies of trust in management science, e-commerce, purchasing, and branding, including much work on human-human and human-company trust, though often technologically mediated.  Thus, they have a relatively weaker showing in the robotics and automation literature review.  On the other hand, some commonly-cited instruments were barely used in practice (e.g., Fluency \cite{Hoffman2013}, SATI \cite{SATI}, CTI \cite{Chien2014}, Schaefer \cite{schaefer2013perception}, the German TiA \cite{Korber2018}), in part reflecting the deep divisions and lack of overlap between even the seemingly closely related fields that study human-machine trust, such as fluency, automation trust, e-commerce, and human-computer interaction.   

\paragraph{Assessing Instrument Quality}
A common finding in our review was the high number of trust instruments were seemingly ad hoc creations.  On the other hand, several other instruments were carefully crafted and painstakingly validated.  Thus, it became clear that our survey should take on a second, important dimension - to compare the quality of the instruments, as well as their availability and utility.  



To assess the current quality of field surveys, we classified them by the quality of their internal reliability and quantitative validity.  As discussed above (Section \ref{sec:rel}), internal reliability can be calculated in many ways.  While the standard is Cronbach's $\alpha$, reporting it alone is insufficient to establish the test's reliability unless its assumptions are met.  This is true of all the internal reliability tests; researchers must account for sample size, missing data, outliers, skew, kurtosis, heterogeneity, and $\tau$-equivalence.  Reporting more than one test also increases confidence that the reliability is accurate.  Thus, we understood a test to have \textit{Complete Reliability} if it reported internal reliability and checked that the data met the test's assumptions. 

For validity, the gold standard is a complete exploratory factor analysis (EFA), including reporting the number of subjects, rotation used, factor loadings, justified choice of factor number, and descriptive statistics - especially including the number of items per factor.  Failing this, however, a confirmatory factor analysis (CFA) then it was considered sufficiently complete if it included Root Means Square Error (RMSEA) and at least two fit statistics such as $\chi^2$, the Standardized Root-Mean Squared Residual (SRMR), the Comparative Fit Index (CFI), or the Tucker-Lewis Index (TLI).  Instruments meeting all appropriate minimally sufficient criteria were classified as reporting \textit{Complete Reliability and Validity}.  If some of these elements were missing, but more effort was made and reported than Cronbach reliability alone, the instrument was categorized as having \textit{Some Reliability and Validity}.  Those that reported no reliability or validity analysis or only Cronbach's internal reliability were classified as \textit{No Reliability or Validity}

It is important to note that carrying out appropriate analyses alone does not mean that the results met any acceptable thresholds, however.  Criteria for assessing the actual quality of the validity and reliability are discussed below in Section \S\ref{sec:goodness}.  Furthermore, while we have done our best to identify the most influential and best-validated instruments currently, this is liable to change. 


Many of the 62 instruments, especially those that measure trust as a single factor, lack reported reliability or validity (Fig. \ref{fig:alluvial}), with 24\% having no validity or reliability analysis of any sort and another 24\% only reporting Cronbach's $\alpha$ for internal reliability.  Only 17\% of studies with a single-factor instrument report any level of both empirical validity and reliability analyses, compared to 54\% of multi-factor instruments.  This is unsurprising, given that multi-factor instruments are subjected to greater scrutiny of dimensional validity.  Even so, 30\% of multi-factor surveys did not report any reliability and validity testing (Fig. \ref{fig:alluvial}).  Despite repeated calls for stronger instrument validation \cite{Gefen2003,Korber2018}, the number of studies utilizing instruments without validation has increased over the last decade.  However, there are currently some promising signs of that trend reversing (see the most recent years in Fig. \ref{fig:hist1}).  

It is worth noting a particular trade-off in types of validity evident throughout our review, lack of survey reuse.  Among those that used or incorporated other surveys, only 17\% used other trust instruments wholesale.  In contrast, the remainder used other previously validated instruments to get at specific individual antecedents or trust-related factors.  Reusing surveys allows for direct comparison and can help validate the reliability and generalizability of an instrument.  On the other hand, creating new surveys can help confirm the nomological/construct validity of both the new survey and the parallel constructs in previous works \cite{boudreau2001validation}.   

The most common of these were instruments that aim to capture \textbf{\emph{Dispositional trust}}, such as either \emph{Faith in Technology} or \emph{Faith in People}.  \emph{Faith in Technology} was most often captured with the Complacency Potential Rating Scale \cite{IdramaniL.SinghRobertMalloy1993} (4 studies) or the Negative Attitudes Towards Robots Scale \cite{nomura2006measurement} (6 studies).  \emph{Faith in People} was measured with instruments like Rotter’s Interpersonal Trust Scale \cite{Rotter_67}, or the instruments of \citeA{moorman1993factors}, \citeA{mcshane2014propensity}, \citeA{wheeless}, and \citeA{anderson1990development}.  The next most common area to try and capture independently are assessments of trust antecedents or correlates such as likeability, bonding, human likeness/anthropomorphism, or affect \cite{Heerink2009,VanDerLaan1997,Merritt2011,kidd2008robots,watson1988development,de2005assessing}.  Other areas measured included \emph{Suspicion} \cite{Lyons2012,selkowitz2017using},\textit{Situation Awareness} \cite{DeVisser2011}, \textit{Mood} \cite{Petersen2018}, \textit{Risk} \cite{Petersen2018,rajaonah2008role,Desai2012}, \textit{Transparency} \cite{Rupp2016}, and \textit{Intention to Work} \cite{Erebak2019}.  Finally, \emph{Workload}, both with the NASA task load index (TLX) and other measures, was the most popular non-trust-related measure (12 studies).

\begin{table*}
    \centering
        \caption{Citation Statistics of the 12 Most Cited, Used, or Impactful Survey Instruments for Capturing Trust across HRI, HAI, HMI, HCI, e-commerce, Technological Acceptance, and Fluency by year, including for citations (cit.) in experimental work (exp.).}
    \resizebox{\columnwidth}{!}{%
    \begin{tabular}{l|c|c|c|c|c|c|c|c||c}
        &&\multicolumn{4}{c|}{\underline{Google Scholar}}&\multicolumn{3}{c||}{\underline{Semantic Scholar}}& \\
    Survey & Year & \begin{tabular}{@{}c@{}}General\\ Citations\end{tabular} & \begin{tabular}{@{}c@{}}Exp.\\ Citations\end{tabular} & \begin{tabular}{@{}c@{}}Gen.\\ Cit./year\end{tabular} & \begin{tabular}{@{}c@{}}Experimental\\ Cit./Year\end{tabular} & \begin{tabular}{@{}c@{}}Background\\ Section\end{tabular} & \begin{tabular}{@{}c@{}}Method\\ Section\end{tabular} & Highly Influential & \begin{tabular}{@{}c@{}}Frequency in \\ Lit. Reviews\end{tabular}\\\hline
      Muir  & 1996 & \cellcolor{red!27}960 & \cellcolor{red!70}351 & \cellcolor{red!7}36 & \cellcolor{red!52}13 & \cellcolor{red!24}613 & \cellcolor{red!20}92 & \cellcolor{red!16}98 & \cellcolor{red!37}8.0\% \\ \hline
      Jian et al. & 2000 & \cellcolor{red!30}1027 & \cellcolor{red!100}473 & \cellcolor{red!10}49 & \cellcolor{red!88}22 & \cellcolor{red!3}78 & \cellcolor{red!20}93 & \cellcolor{red!4}22 & \cellcolor{red!100}21.7\%\\ \hline
      \begin{tabular}{@{}l@{}}Madsen \&\\ Gregor (HCT)\end{tabular} & 2000 & \cellcolor{red!10}293 & \cellcolor{red!26}131 & \cellcolor{red!3}14 & \cellcolor{red!24}6 & \cellcolor{red!3}79 & \cellcolor{red!10}48 & \cellcolor{red!5}34 & \cellcolor{red!9}4.0\% \\ \hline
      Gefen (TAM) & 2003 & \cellcolor{red!100}8832 & \cellcolor{red!23}118 & \cellcolor{red!100}491 & \cellcolor{red!24}6 & \cellcolor{red!100}2353 & \cellcolor{red!100}549 & \cellcolor{red!100}615 & \cellcolor{red!20}4.4\% \\ \hline
      SHAPE (SATI) & 2003 & \cellcolor{red!3}60 & \cellcolor{red!5}24 & \cellcolor{red!1}3 & \cellcolor{red!4}1 & \cellcolor{red!0}6 & \cellcolor{red!1}6 & \cellcolor{red!0}4 & \cellcolor{red!0}0.6\%\\ \hline
      Heerink/UTUAT & \begin{tabular}{@{}c@{}}2009,\\2011\end{tabular} & \cellcolor{red!20}638 & \cellcolor{red!46}231 & \cellcolor{red!11}58 & \cellcolor{red!84}21 & \cellcolor{red!7}174 & \cellcolor{red!25}119 & \cellcolor{red!9}56& \cellcolor{red!13}2.9\%\\ \hline
        McKnight & 2011 & \cellcolor{red!55}5725 & \cellcolor{red!20}100 & \cellcolor{red!60}301 & \cellcolor{red!20}5 & \cellcolor{red!62}1560 & \cellcolor{red!90}428 & \cellcolor{red!78}478 & \cellcolor{red!24}5.1\% \\ \hline
      \begin{tabular}{@{}l@{}}Merritt \\ (Propensity to Trust)\end{tabular} & \begin{tabular}{@{}c@{}}2008,2011,\\2015,2019\end{tabular} & \cellcolor{red!15}510 & \cellcolor{red!48}239 & \cellcolor{red!9}44 & \cellcolor{red!81}21 & \cellcolor{red!8}204 & \cellcolor{red!10}45 & \cellcolor{red!7}46 & \cellcolor{red!13}2.9\% \\ \hline
      Hoffman (Fluency) & 2013 & \cellcolor{red!8}177 & \cellcolor{red!12}62 & \cellcolor{red!4}22 & \cellcolor{red!32}8 & \cellcolor{red!3}62 & \cellcolor{red!10}45 & \cellcolor{red!2}11 & \cellcolor{red!0}0.6\% \\ \hline
      Schaefer  & 2013 & \cellcolor{red!6}121 & \cellcolor{red!11}56 & \cellcolor{red!3}15 & \cellcolor{red!28}7 &\cellcolor{red!2}50 & \cellcolor{red!8}36 & \cellcolor{red!3}16 & \cellcolor{red!3}0.6\%\\ \hline
      Chien (CTI)  & 2014 & \cellcolor{red!2}38 & \cellcolor{red!3}15 & \cellcolor{red!1}5 & \cellcolor{red!8}2 & \cellcolor{red!0}3 & \cellcolor{red!0}0 & \cellcolor{red!0}0 & \cellcolor{red!3}0.6\%\\ \hline
      K\"{o}rber (German TiA)  & 2018 & \cellcolor{red!7}155 & \cellcolor{red!15}75 & \cellcolor{red!10}52 & \cellcolor{red!100}25 & \cellcolor{red!1}11 & \cellcolor{red!4}17 & \cellcolor{red!0}3& \cellcolor{red!3}0.6\%\\ \hline \hline
      \textbf{Total}&&\textbf{18536}&\textbf{1875}&\textbf{1090}&\textbf{137}&\textbf{5193}&\textbf{1488}&\textbf{1383}&\textbf{47.9\%}
    \end{tabular}
    }
    \label{tab:overall_stats}
\end{table*}
 
\paragraph{Terminology Challenges}  Upon first inspection, there appears to be little overlap with how various instruments decompose trust and, thus, how they assess it.  Further probing reveals that a conceptual consensus is emerging behind the terminological differences.  The following section in this chapter will work to provide a unifying language and then apply that language to the selection of papers chosen so that a complete understanding and comparison can be made.  

\begin{table}
\centering
    \caption{Multi-factorial HMI trust surveys that reported sufficient reliability and validity. The number of subjects, number of relevant trust antecedent and factors, and number of inter-factor relationships reported are listed. For factor mappings between the surveys see Tables \ref{tab:termMap1} \& \ref{tab:termMap2}, Appendix A.}
    \resizebox{\columnwidth}{!}{%
\begin{tabular}{l|c|c|c}
\textbf{Survey Name or Authors, if scale unnamed} & \# Subjects & \# Rel. Factors & \# Inter-factor Relations Tested \\\hline
Gefen's TAM with Truat \cite{Gefen2003} & 213 & 8 & 12 \\\hline
Benbasat and Komiak (v1; 2005) \cite{Benbasat2005} & 120 & 7 & 9\\\hline
Komiak and Benbasat (v2; 2006) \cite{Komiak2006} & 100 & 7 & 10 \\\hline
Trust in RA \cite{Wang2008} & 120 & 7 & 6\\\hline
Merritt's collected Trust-related Scales \cite{Merritt2008,merritt2019automation,merritt2011affective} & 255 & 4 & 6 \\\hline
UTUAT with Trust \cite{Heerink2009} & 30 & 8 & 19\\\hline
Trust in Specific Technology \cite{McKnightD2011} & 359 & 7 & 15 \\\hline
Trust in IT Artifacts \cite{Sollner2012} & 284 & 4 & 3 \\\hline
German TiA \cite{Korber2018} & 94 & 4 & 6 \\\hline
Attitudes toward Affective Technology\cite{Freude2019}&303&7&9\\\hline
Hegner, Beldad, \& Brunswick \cite{Hegner2019}&369&5&7\\\hline
Tussyadiah, Zach, \& Wang \cite{Tussyadiah2020}&625&6&5\\\hline
Park \cite{Park2020}&406&7&6\\\hline
TOAST \cite{Wojton2020}&331&2&1\\\hline
SSRIT \cite{Chi2021}&326&4&6\\
    \end{tabular}}
    \label{tab:factor_summary}
\end{table}

\begin{table}[b]
    \centering
        \caption{The number of multi-factorial trust surveys that found each identified factor of significant reliability and validity (N=25). Familiarity includes reputation and recommendation.}
    \begin{tabular}{l|c|c}
    \textbf{Factor} &	\textbf{Number of Surveys} & \textbf{Percent Represented} \\
    \textbf{Dispositional}&&\\
     Faith in Technology&	10 & 40\%\\
     &&\\
     \textbf{Situational}&&\\
     Familiarity  &	6 & 24\%\\
     Situation Normality&	6& 24\%\\
    Emotional Response&	11 & 44\%\\

    &&\\
    Shared Mental Model &	12 & 48\%\\
&&\\
 \textbf{Learned}&&\\    
Structural Trust &9 & 36\%\\
Capability-based Trust &	19 & 76\%\\
Affective Trust &	15 &60\%\\
General Trust &	14 &56\%\\
&&\\
Intention to Use &	12&48\% \\
    \end{tabular}
    \label{tab:factor_summary2}
\end{table}

\section{Mapping and Defining the Factors and Antecedents of Human Machine Trust}
Thus, between the bottom-up analysis of individual items and the clustering of commonalities between the reliable and validated multi-factorial survey instruments,  we ultimately grouped each set of identified categories as cleanly as possible into 10 factors.  A mapping of the reliable and validated factors from the sampled surveys is displayed in Appendix A, Tables \ref{tab:termMap1} and \ref{tab:termMap2}.   The purpose of this is to primarily show that despite the vast diversity in terminology and dozens of varying survey instruments coming from a multiplicity of fields and approaches, the measurement of trust across human-system interaction is converging.

In order to exploit the increased cross-discipline understandability this convergence makes possible, and for greater clarity in this paper, we will now proceed to describe each of these critical factors for understanding human-machine trust.  The following factors are the most commonly found to be distinct, reliable, and have multiple forms of validity (i.e., surface, construct, internal, external).  We do not claim these constructs are perfect or that the terms we have chosen are the best.  However, we believe they capture the emergent consensus among the experimental literature and the collected measures.

One point to consider is which factors should be considered antecedents instead of direct or proper trust factors.  This is a complex question, as different approaches treat trust differently.  For some, \textit{Faith in Technology} would be considered a proper trust factor \cite{IdramaniL.SinghRobertMalloy1993,Hegner2019,Merritt2011}.  Others might consider \textit{Shared Mental Model} one instead \cite{Wojton2020,Rupp2016}.  Herein, because many in the field want to differentiate between antecedents of trust and trust itself, the choice was made to name those factors that are more downstream, per the meta-analysis in Section \ref{sec:meta-anal}, as more direct factors of trust and more upstream ones as antecedents.  However, we acknowledge that this distinction is not sharp and may be of limited utility. 

\subsection{Dispositional Level}
The term \textbf{Dispositional} herein refers to beliefs or attitudes that exist before one is considering or in a particular situation or interaction; we still recognize that this level may be composed of multiple distinct factors.  While there may be many components, they generally fall along three lines: faith in persons, faith in institutions, and faith in technology.  These are not natural dispositions in the same sense that Hoff and Bashir employed the term \cite{Hoff2015} but are generic attitudes.  Some of the oldest multi-item trust measurement instruments were those measuring faith in institutions and persons \cite{Rotter_67,smith2019general}.  However, these surveys have rarely been used in conjunction with HMI research.

On the other hand, faith in technology is relatively common, appearing as a factor within 10/25 reliable and validated multi-factor surveys (see Tables \ref{tab:factor_summary},\ref{tab:factor_summary2}).  Just because \textbf{Dispositional} factors are not in the \textbf{learned trust} layer, that does not mean they, or those in the \textbf{situational level}, are static.  They may get updated but at much slower time scales than \textbf{learned trust}, as previous interactions shape future dispositions, sense of situation normality, familiarity, and future emotional responses.

\subsubsection{Faith in Technology}
This factor is defined by how much faith one has in technology, usually generally, though sometimes in a specific technology.  \emph{Faith in technology} is a form of \textbf{dispositional trust}, that is, it is not particular to the given situation or even technology involved but one's general trust, though it may be specific to a given type of technology or socio-technical interaction (e.g., robots, online shopping, automation).  While this can be thought of as a form of trust, in terms of a given trust interaction, this is more of a trust antecedent or \textit{prior} that may indicate how a person will feel and what they may anticipate upon first encountering a specific system.  We also choose to term it \textit{faith}, in alignment with  \cite{McKnightD2011}, to stress that it may not be based on any evidence or experience.

\emph{Faith in Technology} is also called \emph{Propensity to Trust Machines} \cite{IdramaniL.SinghRobertMalloy1993}, \emph{Trusting Stance}, or simply \emph{Faith} \cite{McKnightD2011}, \emph{Attitude} \cite{Heerink2009,Scopelliti2005,nomura2008prediction}, and \emph{Anxiety} \cite{Heerink2009,nomura2006measurement}.  It is often measured with pre-existing \textbf{\emph{Dispositional trust}} survey instruments aimed at capturing attitudes and anxiety toward robots or automation, such as the Complacency Potential Rating Scale (CPRS) \cite{IdramaniL.SinghRobertMalloy1993}.  One validated standalone variant of this factor that specifically looks at \emph{faith in technology} from the view of high expectations of technology and all-or-none thinking is the Perfect Automation Schema  \cite{merritt2015measuring}.  If the questions are left open about a generic "robot," then this factor may also include the Robot Anxiety Scale (RAS) and the Negative Attitude towards Robots Scale (NARS) \cite{nomura2008prediction}.  However, those scales may correspond more closely to \emph{Emotional response}, below, if measured within a specific situation or interaction.  Given the wide range of facets that the dispositional may capture, one outstanding question is whether this factor is homogeneous.

\subsection{Situational Level}
\textbf{Situational} factors are those that are contextually relevant to the specific situation or interaction but exist prior to the current interaction.  These draw upon previous knowledge or experience with elements of the system at hand (\textit{familiarity}), similar systems (\textit{situation normality}), and what the design of the system or the quality of the interaction evoke (\textit{emotional response}).  These all inform the mental model the user formulates of the system, its goals, capabilities, preferences, and procedures.  \textbf{Situational} factors are not normally considered forms of trust in and of themselves but as their antecedents.  

\subsubsection{Familiarity}
At the heart of the \textbf{\emph{situational}} antecedents stands \textbf{Familiarity}, which is not only about indirect knowledge but also includes direct experience with the vendor, product, or brand.  The first part of \textbf{familiarity} includes indirect acquaintance such as \emph{reputation}, \emph{recommendation}, and \emph{Social Influence} \cite{Korber2018}.  These sub-factors are central to \textbf{Familiarity} in human-human trust \cite{yamagishi2001} and marketing \cite{ha2005effects}, though mostly ignored in HAI/HRI.  It is primarily termed \textbf{familiarity} by most surveys that employ it, except for Scopelliti's somewhat confusingly named category: \textit{Capabilities of Robots} \cite{Scopelliti2005}.   This factor may be overlooked when trust is measured for users interacting with newly developed systems, especially in academic settings, given the technologies' lack of branding and novelty.

\subsubsection{Situation Normality}
In such cases, it may be preferable to focus on \textbf{Situation normality}, which is similar to \textbf{familiarity} but more indirect.  Similar experiences with other technologies or machine behaviors and functionalities create recognizable parallels from which the human's expectations are formed.  It is usually referred to as \textbf{Situation(al) Normality} \cite{Gefen2003,McKnightD2011,Freude2019,Park2020} but sometimes construed under \emph{life-likeness} \cite{Powers2007}, in a limited functional/behavioral sense.  

\subsubsection{Emotional Response}
Defining the appropriate scope of this category and naming it required much thought, and its cohesiveness as a single factor requires further substantiation.  The unique name of this category is based on the increasing acceptance of the term \textit{emotive} to describe the response evoked by robots when studying these factors in relation to trust \cite{schaefer2013perception,Khalid2020,SATI,french2018trust,Kim2020}.  This category spans factors such as \emph{warm} \cite{Lee2015}, \emph{sociable} \cite{Powers2007,Heerink2009}, \emph{pleasant, friendly, likeable} \cite{Merritt2011,Rau2009}, \emph{evoking of comfort}, and \emph{attachment} \cite{Chi2021,Madsen2000}, \emph{engagement} \cite{Park2020}, \emph{seriousness} \cite{Hammer2015}, \emph{dominance} \cite{Powers2007}, and \emph{professionalism} \cite{Skarlatidou2013}.  Related but negatively valenced factors exist may be \textit{anxiety} \cite{nomura2006measurement}, and more general \textit{negative attitudes} \cite{nomura2008prediction} when measured once a situation has been initiated. 

The importance of \textbf{emotional response} has been increasingly acknowledged by HRI, especially in affective robotics \cite{Mann2015,schaefer2013perception,Merritt2011}.  Anthropomorphism likely affects trust through this antecedent, as does the user interface more generally \cite{Deligianis2017,hauslschmid2017supportingtrust,Zanatto2020}.  Note that we chose \textbf{Emotional response} as it focuses on the human's internal psychological assessment and reaction as opposed to the physical or functional properties of the system being interacted with, which better encompasses the factors among the survey instruments.

\subsection{Shared Mental Model}
\textbf{Shared Mental Model} was chosen as an existing term that broadly captures one's perception of a system's understanding of their interaction and the (lack of) confidence that brings about. This category has seen the highest variability due to construct proliferation and poor understanding.  Consequently, our naming schema deviates more from those currently used as this concept than the other categories.  It has been variously called: \emph{Perceived Understandability/Technical Competence} \cite{Madsen2000}, \emph{Intention of Developers+Understanding/Predictability} \cite{Korber2018}, \emph{Human-Robot Interaction} \cite{Scopelliti2005}, \emph{Responsiveness} \cite{Powers2007}, \emph{Process Transparency} \cite{Chien2014}, \emph{Intentionality} \cite{Ullman2014}, \emph{Comfort of Use and Transparency} \cite{Hammer2015}, \emph{Process} \cite{Sollner2012,Park2020}, \emph{Understanding} \cite{Wojton2020}, \emph{Wearable Technology Trust} \cite{Rupp2016}, and \emph{Perceived Ease of Use} \cite{Gefen2003,Benbasat2005,Hegner2019}.  

While similar, the \textbf{Shared Mental Model} can be thought of as a subset of a theory of mind system which infers cognitive states. It specifically is the part which models the trustee's understanding of the trustor's goals, methodological preferences, their relationship and interaction, and what the trustor believes the trustee knows of them personally.  The inferred cognitive states that a shared mental model embedded in a theory of mind support can be termed \emph{Situation Awareness} \cite{Andrews2022}.  \textbf{Shared Mental Models} are the infrastructure that allow \emph{Situation Awareness} but match the factors among the survey literature more clearly, and SMMs themselves are not as directly tied to predictability.  The formation of a \textbf{Shared Mental Model} on the current perceived state of the interaction yields the `highest' level of  \emph{Situation Awareness}, the ability to project how the situation will unfold into the future \cite{Andrews2022}.  However, the SMM itself does not include predictability but serves as an antecedent to it \cite{Andrews2022}, which makes sense as the literature supports that in trust measurement predictability itself is correlated with \textbf{Capability-based Trust}.

While here we categorized it as an antecedent of trust, it seems to exist between situational and learned trust.  The \textit{shared mental model} is initially formed by the \textbf{dispositional} and \textbf{situational} levels but, unlike them, gets quickly and continuously updated during the trust interaction as one learns more.  

\subsection{Learned Trust}
Within the \textbf{\textit{Learned}} level, we are now talking about forms of actual trust and no longer antecedents.  As such, all of these factors directly have \textbf{trust} in their names.  These are factors that continuously are re-assessed in the interaction as it dynamically progresses between the system and the human and appears to be composed of four main factors.

\subsubsection{Structural Trust}
This factor is defined as the belief that the trustee will abide by whatever cultural and behavioral norms, morals, agreements, and laws the trustor expects.  \textbf{Structural Trust} includes the motivations of the trustee to follow such norms, including having a sense of integrity, responsibility, honor, and shame.  This factor has also been termed \emph{Reliance} \cite{IdramaniL.SinghRobertMalloy1993}, in addition to \emph{Credibility-Character} \cite{Rau2009}, \emph{Fairness} \cite{Ullman2014}, \emph{Honesty} \cite{Ullman2014}, and \emph{Integrity} \cite{Benbasat2005}.  This type of trust is closely related to perceived moral agency \cite{Banks2019}, though that is more about the capacity for morality than the trust placed in shared norms.  Hence, why the author of that study depended on \cite{Rempel1985} to measure trust separately.  While an instrument designed for human-human interpersonal trust, \cite{Rempel1985} can be understood as specifically focusing on the \textit{shared mental model}, \textit{structural}, and \textit{affective trust}.

We specifically chose the word \textbf{trust} within the factor name to differentiate from \textbf{assurance}, \textbf{integrity}, \textbf{honesty}, and \textbf{fairness}. It is not \textbf{Structural Assurance}, which only focuses on regulatory and institutional mechanisms \cite{McKnightD2011,Gefen2003, Freude2019,Park2020}, and does not also include beliefs about the trustee's motives or likelihood of compliance with structural mechanisms.  This additional aspect of \textbf{Structural Assurance} integrates a normative framing, while remaining open to both internally and externally driven motives.  In additional to allowing for external motivations, we also choose to call this a type of \textit{trust} instead of \textbf{integrity}, \textbf{honesty}, and \textbf{fairness} as these terms are ambiguous as to whether they refer to beliefs of the trustor or properties of the trustee (i.e., their trustworthiness).   

\subsubsection{Capability-based Trust}
This category, whose name is based on \citeA{Freude2019}, broadly captures whether a system has the resources and capabilities to perform a task (\emph{competence}) \cite{Benbasat2005,Komiak2006,Korber2018}, one's belief in that competence, whether based on experience or faith (\emph{confidence}) \cite{Lee_94,Lee2004,josang2016subjective}, the belief that the system performs the task reliably (\emph{reliance}) \cite{Madsen2000,Korber2018,McKnightD2011,Ullman2014,Tussyadiah2020}, and that the system can recover from errors (\emph{robustness}) \cite{SATI,Madsen2000}  Depending on the system, security, privacy, and accessibility may all be perceived as falling under \textbf{Capability-based Trust}, though they are sometimes more accurately framed as elements of \textbf{Structural Trust}.  Other factors that may capture all or part of these concepts are \textit{usefulness} \cite{Gefen2003,Heerink2009,Heerink2011}, \textit{attitude} \cite{Heerink2009}, \textit{functionality} \cite{McKnightD2011,schaefer2013perception,Tussyadiah2020}, \textit{performance} \cite{Sollner2012,schaefer2013perception,Chien2014,Park2020,Wojton2020}, \textit{fluency} \cite{Hoffman2013}, robot-relative contribution \cite{Hoffman2013}, and \textit{capable} \cite{Ullman2014}.  Some surveys take all trust or trustworthiness as capability-based and term the factor accordingly \cite{IdramaniL.SinghRobertMalloy1993,Merritt2008,Chi2021}.  

Another common but confusing term for this factor is \emph{cognitive or cognition-based trust}\cite{johnson2005cognitive,tussyadiah2018,Komiak2006,schaefer2013perception}, as all trust emerges from cognition.  This usage rested on the assumption that \emph{cognitive trust} was evidence-based and rationally derived as opposed to \textbf{affective trust} arising from blind emotions \cite{lewis1985trust}.  This division is based on the philosophical understanding that one form of trust is irrational because it is tied to emotions.  However, the debates  over the rationality of emotions \cite{pham2007emotion,gubka2022there,haselton2006irrational,solomon1973emotions,de1979rationality} and cognition as computation \cite{van1995might,piccinini2020neurocognitive} are long standing.  Thus, despite less historical weight, we chose the term \textbf{capability-based trust} over \textbf{cognition-based or cognitive trust}.  However, we retained the term \textbf{affective trust}, as explained below.  

\subsubsection{Affective Trust}
\textbf{Affective trust} is trusting that the trustee will support one's goals or actions to achieve those goals.  More generally, it may be understood as how cooperative the trustee is expected to be \cite{gillespie2003measuring,McKnight2001,dunning2014trust}.  At its most narrow, this factor has been termed \emph{Calculative-Based Beliefs} \cite{Gefen2003} and at its broadest and most anthropomorphized, \emph{benevolence} \cite{Mayer1995,Wang2005,Chien2014,Benbasat2005}.   Between the two we find \emph{Purpose} \cite{Chien2014,Sollner2012,Park2020}, \emph{Intention of Developers} \cite{Korber2018}, \emph{Safety} \cite{IdramaniL.SinghRobertMalloy1993}, \emph{Goal} \cite{Hoffman2013},  and \textit{Helpfulness} \cite{McKnightD2011,Sollner2013,Wechsung2013,tussyadiah2018}.  Despite the popularity of Mayer's trust definition \cite{Mayer1995}, we caution against framing this factor as benevolence, which may imply that the robot or machine is kind, caring, and capable of goodwill.

Concepts such as honesty and loyalty may sometimes be construed as either \textbf{Affective} or \textbf{Structural}, or a mix of both.  Their categorization is affected by perceived motivation, such as whether the robot is enabling one's goals because it is aware of them specifically or because it is conforming to societal standards.  Depending on the questions asked, these two constructs may overlap so heavily as to be indistinguishable and be captured under the catch-all term \emph{relation-based trust} \cite{law2021trust}.  Conversely, their opposites may be packed together into an amorphous distrust category that includes deception, trickiness, and underhandedness, like the distrust items in \citeA{jian}.  In general, \emph{distrust} is often meant as the antonym of \textbf{Affective Trust}, which may lead to confusion as to whether it is the opposite of `trust' or whether it is a distinct concept that must be measured independently \cite{Lewicki1998}.

Another limitation of our goal-supportive definition is that it can be at odds with a more dyadic, \textbf{affective trust} that the trustee wants what is best for the trustor.  This can lead to a paternalistic tension between the trustor's self-posited goals and intended means and the trustee's belief in what the trustor should want or how they should act.  As we are primarily interested in goal-focused trust, for now, we simply acknowledge that this can lead to different conceptions of affective trust, especially across cultures where paternalism may be received very differently \cite{Shi2020,Pellegrini2008,Kocak2021,clarke2013trust}.

\subsubsection{General Trust} This category probes trust or trustworthiness directly, though questions can range from focusing on a particular function to the whole system.  Questions here also may ask about trust in those that designed, built, or sold the system.   Other monikers for this category include \emph{Trust in (Service) Robots} \cite{Park2020,Hoffman2013}, \emph{Trust in Automation} \cite{Korber2018}, \emph{Trusting Beliefs in Specific Technologies} \cite{McKnightD2011}, \emph{Perceived Machine Accuracy} \cite{Merritt2011}, \emph{Social Service Robot Interaction Trust} \cite{Chi2021}, and simply \emph{Trust} \cite{Sollner2012,Hegner2019,Freude2019}.

\subsubsection{Intent to Use}
Once we get beyond evaluating all our other internal beliefs, perceived risk, and other environmental and task-related factors, we must finally decide whether to trust.  It is usually called some variation of \emph{Intention} \cite{McKnightD2011,Heerink2009,Benbasat2005,Hegner2019,Park2020,Freude2019}, but also sometimes \emph{Reliance}\cite{Merritt2011} or \emph{Confidence} \cite{Chien2014,IdramaniL.SinghRobertMalloy1993}, which some may find particularly confusing. 

\section{Final Selection of Studies to Analyze}
It is simply beyond the scope of this dissertation to thoroughly analyze for validity, reliability, and terminology used in all 62 instruments.  Instead, a down-selection was performed, it quickly became clear that despite the wide number of instruments only around a dozen were cited frequently or re-utilized among the validated reports, as discussed above in Section \ref{sec:quality}.

As many papers cite from these works, totals in Table \ref{tab:overall_stats} should be taken as estimates of upper bounds on the number of relevant papers.  Here the citations among Method sections, according to Semantic Scholar, best reflect unique experiments, so we can likely put the upper bound on experiments that cite these Top 12 at $ \sim1500 $.  Assuming our sampling method was appropriate, this represents approximately half of the total experiments, so there are likely 3000 relevant experiments, of which our literature review analyzed 173 (5.7\%).  At this sample size, we can have 95\% confidence that the Top 12 really do represent approximately $50\pm7.2\%$ of all survey instruments used in experimental research.  


While there are other contenders for top spots as far as direct experimental usage, these 12 were identified and chosen to analyze herein after assessing their overall impact (as partially illustrated in Table \ref{tab:overall_stats}).  This list is by no means exhaustive, and the choice of 12 was made primarily to scope our study.  However, we believe they represent a substantial sample of the current state-of-the-art, a good range of validated factors, and provide an entry point into assessing how instruments are tested for reliability and validity.

With these definitions of trust and its antecedents in mind, we can now turn to the Top 12 trust instruments and assess their reliability and validity as well as how well they cover these major {\fontfamily{qzc}\selectfont internal} categories that relate to the psychology of trust and its antecedents.  Note that these categories did not include {\fontfamily{qzc}\selectfont external} environmental, or robot-based factors, but were limited to what the human perceives, understands, and expects of the robot.
\section{The Structure of Human-Machine Trust} \label{sec:meta-anal}


The inter-relations between these ten antecedents and trust factors was further refined and validated by closely examining the 15 studies that performed regressions and structural equation modeling.  While not every study tested every factor, the meta-analysis presented in Table 3 reveals some general trends that suggest a common set of relationships, if not an underlying joint structure. 

\begin{table*}[]
    \centering
        \caption{Meta-Analysis of Regression Coefficients.  Adjusted partial correlations (adj.) for publication bias (trim and fill) when appropriate.  Standard Error (SE), hypothesis testing, and significance level (Z-Value), $95\%$ upper and lower confidence intervals (CI).  Also reported are metrics for estimating heterogeneity: Cochrane's Q, $I^2$, $T^2$, and $\tau$ [32]. $*=p<.1, **=p<0.05, ***=p<.01$.  Relations with only one sample have no heterogeneity to report, and their significance (in parenthesis) reflects what was reported by the original authors.  Publication bias is only reported where there are 3+ sources, and heterogeneity did not overly threaten adjustment (Terrin et al., 2003)}
        \resizebox{6in}{!}{
    \begin{tabular}{ll|cccccccccccc}
      \begin{tabular}{@{}c@{}}\textbf{Starting}\\\textbf{Level}\end{tabular}&\textbf{Relation}& \textbf{Sources}&\textbf{partial corr.}&   \begin{tabular}{@{}c@{}}\textbf{partial}\\\textbf{ corr. adj.}\end{tabular}&\textbf{SE}&\textbf{Z-value} &\begin{tabular}{@{}c@{}}\textbf{95\% CI}\\\textbf{ Lower}\end{tabular}&\begin{tabular}{@{}c@{}}\textbf{95\% CI}\\\textbf{ Upper}\end{tabular}&\textbf{Q}&\textbf{p$_Q$}&\textbf{I$^2$}&\textbf{T$^2$}&\textbf{$\tau$}\\
     
     \thickhline
     
     Disp.&\begin{tabular}{@{}l@{}}\textbf{Faith in Technology\protect\MVRightarrow}\\ \textbf{Situational Normality}\end{tabular}  & 2&0.2 & - & 0.04&2.61*** &-0.78 &1.19&4.33&0.037&77\%&.01&0.1  \\\hline
      Disp.&\begin{tabular}{@{}l@{}}\textbf{Faith in Technology \protect\MVRightarrow}\\ \textbf{Shared Mental Model}\end{tabular}  &1& 0.37 & - & 0.05&(***) &0.28&0.46&-&-&-&-&- \\\hline
     Disp.&\begin{tabular}{@{}l@{}}\textbf{Faith in Technology\protect\MVRightarrow}\\ \textbf{Structural Trust}\end{tabular}  &2& 0.23 & - & 0.04&3.82*** &-0.53 &0.99&2.65&0.104&62\%&0.0&0.07  \\\hline
     Disp.&\begin{tabular}{@{}l@{}}\textbf{Faith in Technology\protect\MVRightarrow}\\ \textbf{General Trust}\end{tabular}  &3& 0.25 & 0.30& 0.03&2.77*** &0.19 &0.41&26.14&0.000&92\%&0.03&0.17  \\\hline
      Disp.&\begin{tabular}{@{}l@{}}\textbf{Faith in Technology\protect\MVRightarrow}\\ \textbf{Intent to Use}\end{tabular}  &1& 0.16 & -& 0.05&(***)&0.06 &0.26&-&-&-&-&-  \\ \thickhline
      Sit.&\begin{tabular}{@{}l@{}}\textbf{Emotional Response\protect\MVRightarrow}\\ \textbf{Situational Normality}\end{tabular}  &1& 0.54 & -& 0.13&(**)&0.27 &0.81&-&-&-&-&-  \\\hline
         Sit.&\begin{tabular}{@{}l@{}}\textbf{Emotional Response\protect\MVRightarrow}\\ \textbf{General Trust}\end{tabular}  &1& 0.36 & -& 0.03&(***)&0.29 &0.43&-&-&-&-&-  \\\hline
          Sit.&\begin{tabular}{@{}l@{}}\textbf{Emotional Response\protect\MVRightarrow}\\ \textbf{Intent to Use}\end{tabular}&1  &0.33& - & 0.06& (*)&0.22 &0.44&-&-&-&-&-  \\\thickhline
          Sit.&\begin{tabular}{@{}l@{}}\textbf{Situation Normality\protect\MVRightarrow}\\ \textbf{Faith in Technology}\end{tabular}&1  &0.41& - & 0.16& (*)&0.09 &0.73&-&-&-&-&-  \\\hline
          Sit.&\begin{tabular}{@{}l@{}}\textbf{Situation Normality\protect\MVRightarrow}\\ \textbf{Shared Mental Model}\end{tabular}&1  &0.47& - & 0.05& (**)&0.36 &0.58&-&-&-&-&-  \\\hline
          Sit.&\begin{tabular}{@{}l@{}}\textbf{Situation Normality\protect\MVRightarrow}\\ \textbf{General Trust}\end{tabular}&3  &0.33& - & 0.03& ND &0.33 &0.33&0&1&ND&0&0  \\\thickhline
          Sit.&\begin{tabular}{@{}l@{}}\textbf{Familiarity\MVRightarrow}\\ \textbf{Shared Mental Model}\end{tabular}&2  &0.25& - & 0.05& 54.02*** &0.19 &0.31&0.01&0.933&0\%&0&0  \\\hline
          Sit.&\begin{tabular}{@{}l@{}}\textbf{Familiarity\protect\MVRightarrow}\\ \textbf{Structural Trust}\end{tabular}&1  &0.21& - & 0.50& (*) &0.02 &0.40&-&-&-&-&-  \\\hline Sit.&\begin{tabular}{@{}l@{}}\textbf{Familiarity\protect\MVRightarrow}\\ \textbf{Capability-based Trust}\end{tabular}&2  &0.23& - & 0.07& 11.49*** &-0.02 &0.48&0.08&0.772&0\%&0&0  \\\hline
          Sit.&\begin{tabular}{@{}l@{}}\textbf{Familiarity \protect\MVRightarrow}\\ \textbf{General Trust}\end{tabular}&1 &0.42& - & 0.05& (***) &0.33 &0.52&-&-&-&-&-  \\\hline
          Sit.&\begin{tabular}{@{}l@{}}\textbf{Familiarity \protect\MVRightarrow}\\ \textbf{Intent to Use}\end{tabular}&1  &0.10& - & 0.06& (**) &-0.01 &0.22&-&-&-&-&-  \\\thickhline
          Learn.&\begin{tabular}{@{}l@{}}\textbf{Shared Mental Model\protect\MVRightarrow}\\ \textbf{Capability-based Trust}\end{tabular}&4  &0.41& 0.42 & 0.03& 7.99*** &0.34 &0.50&10.64&0.014&71\%&0.01&0.08  \\\hline
          Learn.&\begin{tabular}{@{}l@{}}\textbf{Shared Mental Model\protect\MVRightarrow}\\ \textbf{General Trust}\end{tabular}&7  &0.47& 0.52 & 0.02& 6.65*** &0.47 &0.57&86.11&0.00&93\%&0.04&0.19  \\\thickhline
          Learn.&\begin{tabular}{@{}l@{}}\textbf{Structural Trust\protect\MVRightarrow}\\ \textbf{General Trust}\end{tabular}&5  &0.40& - & 0.02& 19.45*** &0.34 &0.46&3.26&0.515&0\%&0&0  \\\hline
          Learn.&\begin{tabular}{@{}l@{}}\textbf{Structural Trust\protect\MVRightarrow}\\ \textbf{Intent to Use}\end{tabular}&1  &0.12& - & 0.10& (*) &-0.02 &0.38&-&-&-&-&-  \\\thickhline
          Learn.&\begin{tabular}{@{}l@{}}\textbf{Capability-based Trust\protect\MVRightarrow}\\ \textbf{Shared Mental Model}\end{tabular}&1  &0.47& - & 0.15& (**) &0.16 &0.78&-&-&-&-&-  \\\hline
             Learn.&\begin{tabular}{@{}l@{}}\textbf{Capability-based Trust\protect\MVRightarrow}\\ \textbf{General Trust}\end{tabular}&7  &0.47& - & 0.01& 5.56** &0.26 &0.67&224.74&0&97\%&0.05&0.21  \\\hline
             Learn.&\begin{tabular}{@{}l@{}}\textbf{Capability-based Trust\protect\MVRightarrow}\\ \textbf{Intent to Use}\end{tabular}&3  &0.28& 0.33 & 0.05& 3.32*** &0.12 &0.53&4.83&0.09&58\%&0.01&0.12\\\thickhline
            Learn.& \begin{tabular}{@{}l@{}}\textbf{Affective Trust\protect\MVRightarrow}\\ \textbf{Structural Trust}\end{tabular}&1  &0.43& - & 0.08& (**) &0.27&0.59&-&-&-&-&-\\\hline
              Learn.&\begin{tabular}{@{}l@{}}\textbf{Affective Trust\protect\MVRightarrow}\\ \textbf{Capability-based Trust}\end{tabular}&2  &0.66& - & 0.04& 2.92*** &0.27&0.59&24.09&0&96\%&0.10&0.31\\\hline
              Learn.&\begin{tabular}{@{}l@{}}\textbf{Affective Trust\protect\MVRightarrow}\\ \textbf{General Trust}\end{tabular}&8  &0.37& - & 0.01& 4.40*** &0.17&0.57&341.54&0&98\%&0.080&0.28\\\thickhline
             
             Learn.&\begin{tabular}{@{}l@{}}\textbf{General Trust\protect\MVRightarrow}\\ \textbf{Faith in Technology}\end{tabular}&1  &0.36& - & 0.17& (*) &0.02&0.70&-&-&-&-&-\\\hline
             
             Learn.&\begin{tabular}{@{}l@{}}\textbf{General Trust\protect\MVRightarrow}\\ \textbf{Capability-Based Trust}\end{tabular}&1  &0.26& - & 0.06& (**) &0.02&0.70&-&-&-&-&-\\\hline
             
             Learn.&\begin{tabular}{@{}l@{}}\textbf{General Trust\protect\MVRightarrow}\\ \textbf{Intent to Use}\end{tabular}&7  &0.55& 0.63 & 0.01& 8.53*** &0.60&0.66&110.83&0&95\%&0.03&0.16\\\thickhline
    \end{tabular}
    }
    \label{tab:meta_anal}
\end{table*}


As can be seen in Table \ref{tab:meta_anal} at the \textbf{\emph{Dispositional trust}} level,  \textbf{Faith in Technology} has the strongest effect on the \textbf{Shared Mental Model}, followed by \textbf{Structural} and \textbf{Capability-based Trust}, \textbf{General Trust}, \textbf{Situational Normality}, and finally, and most weakly, on \textbf{Intention to Use}.  Its effect weakens over the interaction \cite{Merritt2008} and over repeated interactions \cite{Sollner2013} as one gets used to a specific technology and better calibrates their trust.  Thinking about trust as a Bayesian expectation (such as in \citeA{josang2016subjective}), \textbf{Faith in Technology} essentially serves as an initial \emph{prior} belief whose contribution diminishes over time.  Therefore, \textbf{Faith in Technology} is most crucial at the initial stages of trust formation, when potential users decide whether to accept and try out new technology.  \textbf{Faith in Technology} gets updated through feedback as both \textbf{General Trust}, and the \textbf{Shared Mental Model} more generally, are calibrated \cite{Heerink2009,Hoff2015}.  

The factors within the \textbf{\emph{Situational trust}} level, like those in the \textbf{\emph{Dispositional}} level, show larger effects early on that diminish overuse \cite{Merritt2008}.  Likeability can even account entirely for early usage of a system until trust is calibrated \cite{Merritt2011}.  \textbf{Emotional Response} is significantly correlated (partial corr: 0.54**, SE: 0.13) with \textbf{Situation Normality} \cite{Heerink2009}.  This relationship, perhaps, is best exemplified by anthropomorphism, which increases naturalness, likeability, and comfort that all directly impact \textbf{Situation Normality} \cite{Chi2021}.  Likewise, there is some evidence that \textbf{Situation Normality} increases comfort and attachment \cite{Chi2021}.  \textbf{Situation Normality} and  \textbf{Emotional Response} both appear to be significant and fairly strong antecedents (partial corr: 0.33-0.47) to more `downstream' factors such as the \textbf{Shared Mental Model}, \textbf{General Trust}, and \textbf{Intent to Use}.  \textbf{Familiarity}, on the other hand, seems somewhat weaker (partial corr: 0.1-0.42), perhaps some of its explanatory power being divided with \textbf{Situation Normality}.  It appears plausible from the SEM models that \textbf{Familiarity} primarily affects the \textbf{Shared Mental Model}, and through that, the sub-factors of trust (\textbf{Affective, Structural, Capability-based}), and eventually \textbf{Intent to Use}.  Few experiments \cite{Heerink2009,Chi2021} have separately tested the sub-factors of \emph{reputation} and \emph{recommendation}, for which there are mixed results from the human-human side \cite{yamagishi2001}. Because the above strength of \textbf{\emph{Situational Trust}} factors weaken over experience, their effect on \textbf{Intent to Use} and actual use may be underestimated or exhibit higher variance, depending on the point in time at which they are measured.

The \textbf{Shared Mental Model} sits at the foundation of \textbf{\emph{Learned trust}} and is crucial in supporting \textbf{Capability-based} (partial corr adj:0.42, SE 0.03) and \textbf{General Trust} (partial corr adj:0.52, SE 0.02).  Like other \textbf{\emph{Learned trust}} factors, the \textbf{Shared Mental Model} gets calibrated over time and is correlated with transparency and the ability to monitor (and thus is also affected by workload \cite{Razin2021a}).  At the crux of the \textbf{Shared Mental Model} is recognizing the goals of the robot \cite{Hoffman2013,tomlinson2020revisiting} and the developers' intentions \cite{Korber2018}, such that one can determine if the robot or machine shares the human's goals or at least is cooperative vs. competitive \cite{castelfranchi2010trust}.  Likewise, the \textbf{Shared Mental Model} is equally necessary for determining whether another will follow laws or norms and hence warrant \textbf{Structural Trust} \cite{Razin2020}.


\textbf{Affective trust} is the foundation of the main trust factors, indicating expected cooperativeness and goal adoption.  Expected cooperativeness vs. competitiveness is critical to understanding whether the potential actors and situation call for trust \cite{Razin2021b}, while goal adoption is the very purpose of task-focused interpersonal trust \cite{castelfranchi2010trust}.  \textbf{Affective Trust} correlates with both \textbf{Structural} (partial corr: 0.43, SE: 0.08) and \textbf{Capability-based Trust} (partial corr: 0.66, SE: 0.04), as well as \textbf{General Trust} (partial corr: 0.37, SE: 0.01).  The extent literature remains unclear as to how \textbf{\emph{learned}} \textbf{Affective Trust} is formed or what its antecedents are, though it may be directly shaped \textbf{\emph{Dispositionally}} \cite{schaefer2013perception,tussyadiah2018}.

Despite its relative lack of attention in the HRI/HAI trust literature, the effect of \textbf{Structural Trust} on \textbf{General Trust} has proven surprisingly strong (partial corr:0.40, SE: 0.02).  Its strength may be an indication that, indeed, ethics and integrity are relevant to specific robotics applications in contradiction to earlier suppositions \cite{Tegmark}.  While \textbf{Structural Trust}'s importance may be application specific, the minor variance across experiments in its effective strength indicates otherwise.  However, \textbf{Structural Trust} may have a weaker effect on \textbf{Intent to Use} (partial corr: 0.12, SE: 0.10) because it gets over-ridden by other factors (e.g., \textbf{Capability-based Trust}, \emph{workload}) \cite{Tegmark}.

\textbf{Capability-based Trust} is usually the go-to in HAI/HRI/HCI; however, these results indicate that focusing on it alone may be a mistake.  \textbf{Capability-based Trust} is only marginally more strongly correlated with \textbf{General Trust} (partial corr: 0.47, SE: 0.01) than \textbf{Structural} (partial corr: 0.40, SE: 0.02) or \textbf{Affective Trust} (partial corr: 0.37, SE: 0.01).  On the other hand, \textbf{Capability-based Trust} has a much stronger direct effect on \textbf{Intent to Use} (partial corr adj: 0.33, SE: 0.05).  It also becomes increasingly important over time, replacing the effects of \textbf{\emph{Dispositional}} and \textbf{\emph{Situational Trust}} \cite{Merritt2008}.  Many instruments found that \textbf{Capability-based Trust} encompassed two distinct sub-factors, often called \emph{Performance} and \emph{Reliability}.  The former can be defined as one's belief in another's ability, while the latter is their confidence assessment of that belief.  For example, I may think that a machine is accurate 80\% of the time, but I may only be 30\% sure that I am correct.  How these beliefs may be combined into a singular assessment is another matter (two ways forward may be Dempster-Shafer Theory or Subjective Logic \cite{josang2016subjective}).  A similar dichotomy between expectation and confidence may be paralleled in \textbf{Structural Trust} with the sub-groupings of \emph{Ethical} and \emph{Sincere}, respectively  \cite{law2021trust}.  While no survey has yet demonstrated a similar dichotomy in \textbf{Affective Trust}, the linguistic cluster analysis in \cite{jian} suggests that \emph{Wariness} may serve as the inverse of \emph{confidence} in \textbf{Affective Trust}.   

While \textbf{General Trust} does not fully determine \textbf{Intent to Use}, it is strongly correlated with it (partial corr adj: 0.63, SE: 0.01).  While strong, this amount of correlation reinforces that \textbf{Intent to Use} is not just a matter of trusting beliefs but is affected by a range of external factors, such as cognitive workload and perceived risk \cite{Desai2012}.

\section{Structure of HRI Trust - An Emergent Model }

\begin{figure*}
    \centering
    \includegraphics[width=0.9\textwidth]{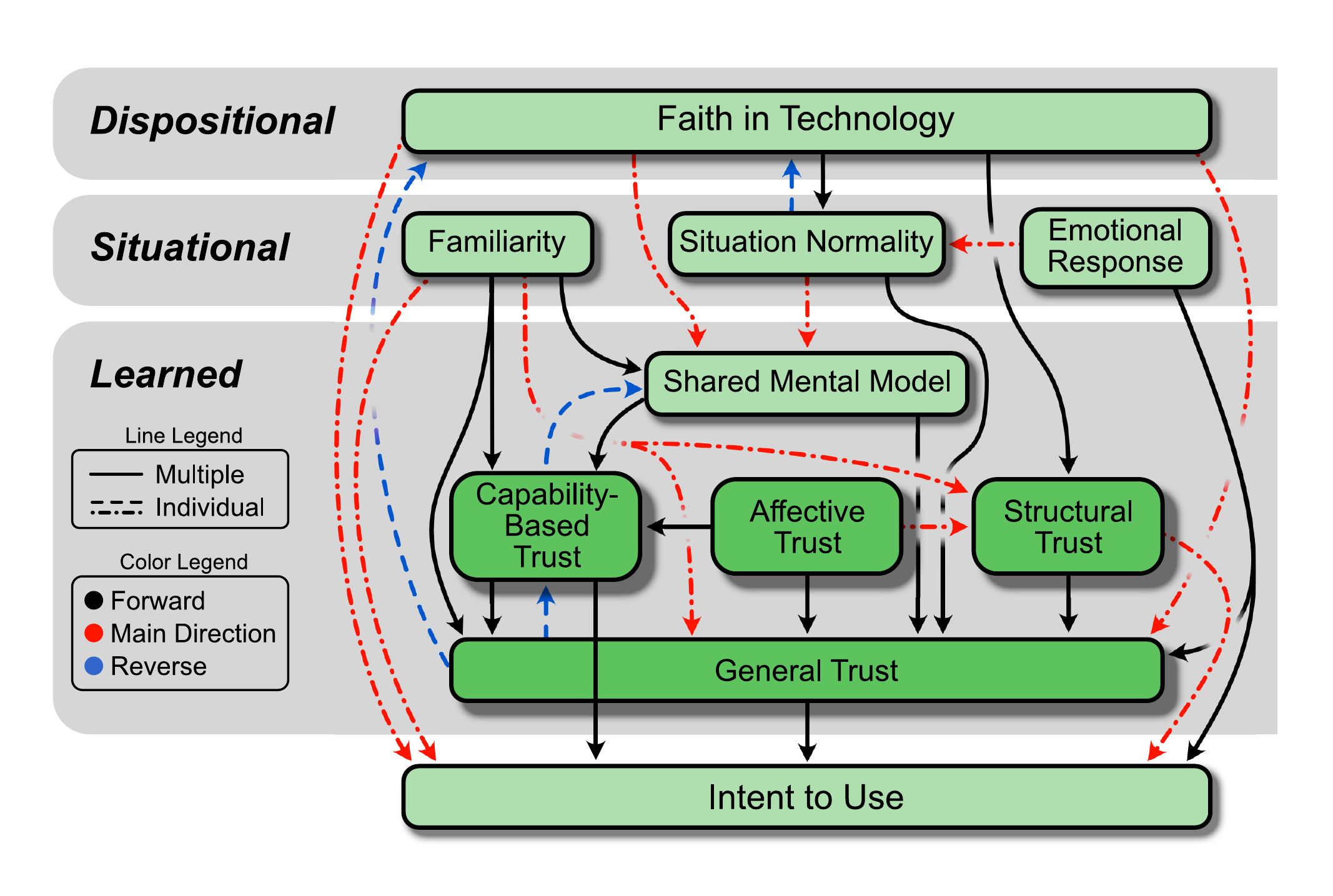}
    \caption{Summary of relations between trust antecedents and factors that emerge from the meta-analysis. Black arrows indicate directed connections established by multiple independent works.  Red and blue lines were only established by individual works.  Red and black indicate connections from dispositional to situational to learned,  and blue arrows are in the reverse direction. }
    \label{fig:caus_graph}
\end{figure*}

The meta-analysis reveals a convergent picture of HAI/HRI Trust emerging out of the efforts of many researchers.  Figure~\ref{fig:caus_graph} provides a graphical summary of this emergent model of trust.  As can be seen, the trust factors and antecedents can be fairly neatly grouped into three layers, substantiating the early findings of \citeA{Hoff2015}.  The overall direction of plausible causality and strong inter-survey agreement has revealed a tentative picture of trust that is, in many ways, intuitive. \emph{A priori} \textbf{\emph{Dispositional}} expectations are first filtered through contextual, \textbf{\emph{Situational}} trust antecedents such as one`s \textbf{Familiarity} with the system, its similarity to previous systems (\textbf{Situational Normality}), and one's impression of the system (\textbf{Emotional Response}). These antecedents all feed into one's \textbf{Shared Mental Model} of the system, which then supports three different types of expectations - whether the system shares the user's goal (\textbf{Affective}), whether it is \textbf{Capable} of achieving that goal, and whether it will follow expected norms or laws when trying to achieve that goal (\textbf{Structural}).  They all contribute to a \textbf{General} assessment of trust, which in turn influences their \textbf{Intent to Use} the system.  Additional studies have further bolstered this model, especially regarding the dynamics of trust over time.  Throughout the interaction, the higher-level factors play decreasingly important roles, as the \textbf{Shared Mental Model} gets updated and the \textit{learned} trust factors  calibrate more closely to {\fontfamily{qzc}\selectfont external} circumstances.

\subsection{Are the differences between Affective and Structural Trust substantial enough to merit separate categories?}

The best way to conceive of \textbf{Structural Trust} is to go back to Luhmann \cite{luhmann}, the originator of the whole science of trust.  \textbf{Structural}, or what he called societal trust, is the abstraction of \textbf{Affective}, or personal, trust.  \textbf{Affective Trust} is founded in the dyadic trust relationship, is rooted in personal knowledge, and is realized through individual attachment, bond, loyalty, and love.  In contrast, \textbf{Structural Trust} universalized this in a world of strangers. 

Thus, while the two are related and both fall under \emph{relation-based trust} \cite{law2021trust}, there are also fundamental differences between \textbf{Structural} and \textbf{Affective Trust}.  \textbf{Affective Trust} is driven by the particular, recognizing the individual's specific goals and supporting them.  In human-human trust, this can be out of benevolence, loyalty, love, or attachment.  However, we have contended that these sources for \textbf{Affective Trust} are not essential, especially in the human-machine context \cite{Razin2021b}.  Instead, we can understand it as arising from inter-agent expectations of commitment and cooperation to achieve shared goals.  On the other hand, \textbf{Structural Trust} is about fairness and equality of treatment.  In humans, it may be driven by fear of sanction or internally-driven honor or integrity to uphold universal norms \cite{dunning2014trust}, and in general, is about how power dynamics encourage compliance around how goals are achieved \cite{Razin2021b}.

That being said, as one factor is derived from the other, the difference between them may not always be so clear, as is illustrated by the correlation between \textbf{Affective} and \textbf{Structural Trust}.  This is because it can be unclear whether honesty, loyalty, and respect should be situated as \textbf{Affective} or \textbf{Structural}. How one differentiates these concepts can depend on cultural understanding, personal motivations, ethical framing, and whether they are attributed to internal or external motivations \cite{dunning2014trust,Chien2014,malle2021multidimensional}.

A common question arises as to whether \textbf{Affective} and \textbf{Structural Trust} are both worth measuring, as many may not see how they are relevant to a specific robot pick-and-place task in a lab, for example.  That the robot could be nefarious and have competitive goals (\textbf{Affective Trust}) or not follow the rules (\textbf{Structural Trust}) is often not even considered).  Asking about deception can, in fact, elicit suspicion in both of these categories, an issue to which we will return.  \textbf{Affective Trust} is likely always at play, even if it is only in the narrowest \emph{calculative-based} sense, i.e., \cite{Gefen2003}.  Ignoring \textbf{Structural Trust} in the lab presents a severe threat to etiological validity where it may have clear effects on behavior and beliefs in the real world when it comes to branding; attitudes toward tech companies, government, and industry regulations; and insurance liability.  Furthermore, AIs and robots are assumed to be more fair and ethical than humans \cite{starke2021fairness}, demonstrating potential bias in \textbf{Structural Trust}.  


Another reason to differentiate between \textbf{Affective}, \textbf{Structural}, and \textbf{Capability-based} trust is that antecedent factors have different effects on them.  For instance, \textbf{Structural Trust} is directly affected by \textbf{\emph{Dispositional}} factors, whereas \textbf{Affective Trust} does not seem to be.   We rely on structural assurance to protect us from strangers and new technologies, linking \textbf{Faith in Technology} and \textbf{Structural Trust}. Another example of the differential effect of antecedents was found in our own experiments \cite{razin2019}, where \textbf{Familiarity} affected \textbf{Capability-based Trust} but not \textbf{Structural}. By dividing out the factors that compose trust, we can better understand which antecedents affect which aspects of our expectations and better understand the {\fontfamily{qzc}\selectfont internal} dynamics of trust.



\subsection{Comparison with Human-Human Trust Measurement}
Unlike HRI/HAI trust, early works on human-human trust were entirely focused on \textbf{\emph{Dispositional trust}} measures \cite{Rotter_67, gillespie2003measuring, mcevily2011measuring}.  Measures for inter-personal trust started developing in many professional contexts, from organizational trust between employees and managers, between firms, and within business networks \cite{Mayer1995,bhattacherjee2002individual,schoorman2007integrative,johnson2005cognitive}.  A second major approach developed that studied trust in friendships and intimate relationships \cite{gottman2011science,bukowski1994measuring}.  Building off these approaches, other veins developed in the healthcare community between patients and doctors \cite{thom2004measuring,anderson1990development}, as well as trust in media \cite{matthes2008content}, and trust in strangers \cite{ermisch2009measuring}.  

While HRI trust rarely explored \emph{Faith in People} or \emph{Faith in Technology Companies}, it did occasionally employ or adapt questions from human-human works \cite<e.g.,>{Rotter_67,wheeless,Mayer1995,Hoffman2013,Mann2015}, though generally did not incorporate them into their larger model or show that they correlated with other antecedents or trust factors.  The influence and trust definition of \citeA{Mayer1995}, who primarily worked in organizational trust, is hard to overstate.  Human-human trust work should also be credited for dividing out \textbf{Affective} from \textbf{Capability-based Trust} \cite{lewis1985trust}, as well as \textbf{Structural Assurance} from \textbf{Situational Normality} \cite{McKnight2001}. 

An essential point of comparison to this study is \citeA{mcevily2011measuring}, who performed a similar review for human-human trust scales.  Interestingly, they identified a similar number of trust scales in human-human trust with a similar pattern of development, where 60\% had created their own \emph{ad hoc} measures, and 40\% re-used a previously validated instrument.  They also discussed the reliability and construct validity patterns over time.  During their review, they found that even fewer human-human scales had reported empirical validity measures than we found for HRI/HAI/HCI scales.  Only 22\% of human-human scales considered multi-factorial trust as opposed to single uni-dimensional items.

The dimensions reported fall closely in line with our own, as seen in Table \ref{tab:hu-hum}. 

\begin{table}
    \centering
    \caption{Operationalized factors of human-human trust \protect\cite{mcevily2011measuring}}
    \begin{tabular}{l|c}
      Factor&\# of times operationalized\\\thickhline
       Capability-based Trust  & 58  \\\hline
       Affective Trust     &  40 \\\hline
       Structural Trust & 22 \\\hline
       General Trust & 9 \\\hline
       Familiarity & 2\\\hline
       Emotional Response & 3\\\hline
       Faith in People & 4\\\hline
       Willingness to Risk & 4 \\\hline
       \begin{tabular}{@{}l@{}}\textbf{Openness, Availability,}\\ \textbf{Receptivity, Forbearance}\end{tabular} & 13\\ 
    \end{tabular}
    \label{tab:hu-hum}
\end{table}

Beyond analogous factors, the similarity in structure is more than apparent.  Thus, trust appears to be exapted; trust mechanisms that developed over millennia for humans and animals are co-opted for robot interaction.  This resemblance lends support to the model of Computers-as-Social actors (CASA) \cite{nass1997computers}, though the internal weights or expectations assigned to individual factors may indeed differ \cite{deVisser2016almost}.

There was only one major grouping that our model did not capture, which was how available, open, and receptive the potential trustees and trustors were.  Likely, this grouping is closely tied to the importance of \emph{Agreeableness} in trust from Personality research \cite{freitag2016personality}, as well as our factor of \textbf{Emotional Response}.  However, it seems distinct enough within human-human trust to merit its own factor and a potential focus of future work in HRI.  On the other hand, \textbf{Shared Mental Models} were entirely missing in \citeA{mcevily2011measuring}.  However, while a decade ago, human-human trust may have lacked such measures, the importance of this factor in human-human trust is now being recognized \cite<e.g.,>{mccomb2017evaluation}.  While many have considered the differences between human-human and human-robot trust \cite{Madhavan2007,Lyons2012,deVisser2016almost,alarcon2021role}, the relevant factors and their measurement are highly similar, and we hope that works like this one will continue to bridge the gap.

\subsection{Limitations and Outstanding Issues}
As this is a meta-analysis where not every survey covered every factor, many factors were not assessed for their potential as mediating factors.  Furthermore, of the 29 relationships examined, 15 are only supported by a single source in our sample, 5 by two sources, and only 9 have three or more sources.  Of those 9, 7 exhibit high heterogeneity, making it challenging to identify which values are outliers or to detect certain biases.  There are also 5 relationships whose lower confidence interval bounds cross 0, which threatens their significance level.  Additionally, there are two relationships where the standard error is nearly as large as, if not larger(!) than, the partial correlation.   Finally, there are three relationships (see blue dashed arrows in Fig. \ref{fig:caus_graph}) where the plausible direction of causality is reversed in one source compared to all other surveys.  Many of these relationships require further repeated testing with reliable and valid instruments over large samples to clarify their significance and plausible direction.

Another challenge is correctly understanding what structural equation models (SEMs) imply and do not imply with regard to causality.  There are many misunderstandings about SEMs, and it should be understood that SEMs do not prove causality.  However, \emph{if} the assumptions of SEMs and statistical measures of fit prove sufficient, then evidence for the plausibility of causality \cite{bollen2013eight} is gained.  While we combine SEM and regression results in the meta-analysis, it is also worthwhile to clarify that SEM and regression are \emph{not} the same, as SEM incorporates causally-relevant assumptions \cite{bollen2013eight}.  This does not preclude combining them in a meta-analysis, but we can no longer assume those assumptions hold.

Another issue can arise with modeling; too few questions may be asked to establish individual factors.  This issue caused us to not include 3 reliable and valid instruments in our meta-analysis.  For example, single \textbf{Affective trust} items focusing on cooperativeness, shared goals, and benevolence often cluster with \textbf{Calculative-based} factors due to their high correlation (partial corr:0.66, SE 0.04) when an insufficient number of questions on \textbf{Affective trust} are asked (e.g., \cite{schaefer2013perception}).  This merger is especially true for surveys that treat trust as a single factor \citeA<e.g.,>{Rupp2016}'s use of \citeA{jian}.

\begin{figure*}
    \centering
    \includegraphics[width=\textwidth]{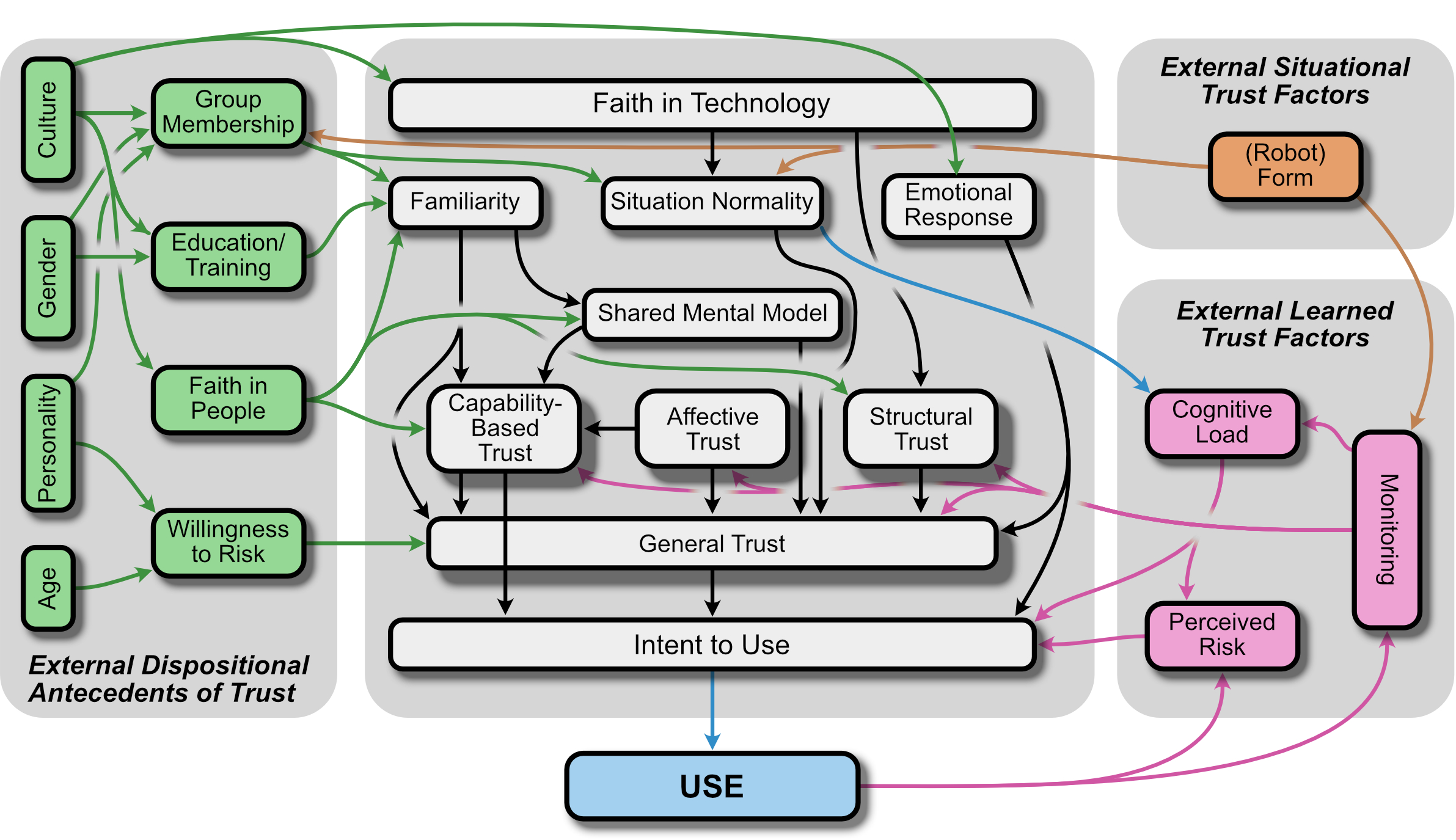}
    \caption{The emergent trust model in relation to significant external factors}
    \label{fig:expanded_trust}
\end{figure*}
\section{Analysis of the Trust Instruments}\label{sec:goodness}
\subsection{Down Selection to a Representative Dozen}

\paragraph{Terminology Challenges.}  Upon first inspection, there appears to be almost no overlap with how the various instruments decompose trust, the terms they use for various factors and antecedents, and how they assess trust.  This may contribute to the continued popularity of the uni- and bi-dimensional trust instruments.  Upon closer inspection, a consensus is emerging, however much it is obscured.  The next section in this chapter will work to provide a unifying language and then apply that language to the selection of papers chosen so that a complete understanding and comparison can be made.

\paragraph{Final Selection of Studies to Analyze}
It is simply beyond the scope of any paper to thoroughly analyze all 62 instruments for validity, reliability, and terminology.  Instead, a downselect was made to select between 10-20 studies to investigate.  

 As many papers cite from these works, totals in Table \ref{tab:overall_stats} should be taken as estimates of upper bounds on the number of relevant documents.  Here the citations among Method sections according to Semantic Scholar may best reflect unique experiments, so we can likely put the upper bound on experiments that cite these Top 12 at $\sim1500$.  Assuming our sampling method was appropriate, this represents approximately half of the total experiments, so there are likely 3000 relevant experiments, of which our literature review analyzed 173 (5.7\%).  At this sample size, we can have 95\% confidence that the Top 12 really do represent approximately $50\pm7.2\%$ of all survey instruments used in experimental research.  

While there are other contenders for top spots as far as direct experimental usage, these 12 were identified and chosen to analyze herein after assessing their overall impact (as partially illustrated in Table \ref{tab:overall_stats}).  This list is by no means exhaustive, and the choice of 12 was made to scope our study.  Still, we believe they represent a robust sample of the current state-of-the-art, a good range of validated factors and provide an entry point into assessing how instruments are tested for reliability and validity.

\subsection{Instrument: Human-Machine Interaction (HMI), (Muir, 1987, 1994, 1996)}
\paragraph{Description.}
The earliest Human-Machine Interaction (HMI) trust scale, Muir \cite{Muir1987}, is primarily focused on measuring performance-based \textbf{\emph{Learned Trust}}.  It consists of nine questions assessed after an interaction occurs.  Muir proposed three main sub-constructs of trust: \textit{faith, dependability}, and \textit{predictability} but also included overall trust in the system and trust in specific elements of the system. \textit{Reliability over time} was grouped with \textit{predictability}, and \textit{competence} and \textit{responsibility} with \textit{dependability}. \textit{Faith} here meant faith in future robustness and not \textbf{Faith in Technology}.  Ultimately, Muir's short survey is a measure of \textbf{Capability-Based Trust}.  

\paragraph{Construct Mapping.}
Being the first published scale for HMI trust, Muir saw widespread use early on.  Muir picked her scale items based on a combination of more general human-human interaction theories \cite{barber1983logic,Rempel1985}, starting with their trust definitions, developing them, and proposing a basic mathematics of trust and trying to substantiate it.

A few key findings concerning the Muir scale are \textit{1)} that it detected that \textbf{Familiarity} and \textbf{Faith in Technology} play more significant roles earlier in an interaction than later \cite{Muir1987,Desai2012}, \textit{2)} that the scale is less sensitive to overall trust changes during an interaction \cite{Desai2012}, \textit{3)} that the factors relating to trust in Muir's scale correlate more with error rate than uncertainty \cite{uggirala2004measurement} (highlighting the scale's focus on  \textbf{Capability-based Trust} vs. \textbf{General} or \textbf{Affective Trust}), and \textit{4)} that their \textit{competence} factor was the only one that correlated with uncertainty.  This single correlation is surprising as one would have surmised both \textit{Faith over Time} and \textit{Predictability} would have been correlated as well, and brings up some fundamental questions as to the scale's content validity \cite{uggirala2004measurement}.  On the other hand, the scale has been shown to correlate strongly with Jian's survey \cite{jian,Desai2012}, though what this may mean is complicated, as discussed in the next section.

Furthermore, deriving meaning from the internal correlations among Muir's items can be messy, as single questions inherently have higher variance and do poorly at examining sub-factor constructs \cite{Korber2018}.  Adding to the variance was the small numbers of participants in her experiments, ranging from 6-12 subjects each \cite{Lee_94,muir1996trust}.  When used to measure between-group trust measures in a driving task, the scale captured trust magnitude and calibration \cite{xiong2012use}. 

\paragraph{Validity \& Reliability.}
Despite its history of usage and popularity, this scale has rarely been tested quantitatively for reliability or validity.  The closest is a refined six-item version that showed Cronbach's $\alpha=0.87-0.91$ \cite{Merritt2011}.  Overall, Muir's scale is short, making it attractive for quickly capturing trust at various times in experiments that can handle brief interruptions.  Its widespread use and citation can partly be attributed to age.  High repeated usage implies cross-application reliability.  However, its lack of internal reliability and validation testing, as well as additional concerns that have arisen experimentally, lead us not to recommend the scale any longer.  It has been superseded by better instruments, including single-factor scales, with more extensive validation and reliability.

\subsection{Instrument: Trust in Automated Systems, (Jian et al., 2000)}
\paragraph{Description.}
\citeA{jian}'s scale of Trust in Automated Systems (TAS or TASS\footnote{Sometimes referred to as Empirically Derived (ED), especially to differentiate it from the Technological Adoptiveness Scale (TAS)\cite{halpert2008technological}}) presently is the most widely used trust scale, and thus is the most important to fully analyze.
TAS aims to capture general trust, as well as to compare human-human and human-automation trust.  The method it employs to understand the concepts underlying trust did not come from a larger model or theory of trust.  Instead, it used linguistic association and distance of words relating to trust.  Ultimately, the 12-item single-factor trust scale that emerged has only seen use in the HAI/HRI/HCI communities rather than in human-human trust applications. 

\paragraph{Construct Mapping.}
TAS consists of 30 words (reduced from an initial set of 176) clustered and mapped into a 12-item scale based on conceptual distances found during the experiment.  Only the most strongly affective words were retained, leading to the dismissal of many more neutral or mixed-effect words related to \textbf{Capability-based Trust} and \textit{certainty}, such as \textit{competence, cooperation, unerring, certainty, definite, absolute}, and \textit{predictability}. Two other words of note that were also dropped were \textit{commit} and \textit{stable}.

Due to its preference for word effect, TAS is almost entirely focused only on \textbf{Affective Trust} and \textbf{Structural Trust} with nods to \textbf{Familiarity} and \textbf{General Trust}.  However, in practice, TAS is often used instead as a measure of \textbf{Capability-based Trust}, and it correlates well with other \textbf{Capability-based Trust} scales such as \cite{Muir1987, Merritt2011}.  This is surprising as the designers of TAS dropped linguistic terms relating to capability, as they tended to have neutral or mixed valences.  Their exclusion of capability, however, is seemingly obscured by scale items such as `reliability,' 'dependability,' `integrity,' and `confidence.'  These terms may have been misconstrued or misunderstood as these words have multiple meanings, complicating their use in trust measurement.

Thus, while TAS can be said to measure trust, it is harder to be sure whether its underlying factors have much construct validity, though they have concept validity (see Section \ref{sec:val}).  This dissembling and disassembling of trust factors has understandably led to a mismatch between the mental models of the scale designers, researchers, and experiment participants, which reflects a fundamental weakness of this scale.


\paragraph{Validity \& Reliability.}
Originally, TAS was developed through three rounds, with the cluster analysis being based on a sample of just 30 students, though now it has been used by hundreds if not thousands in subsequent experiments \cite{gutzwiller2019positive}.  TAS generally has been found to have strong reliability ($\alpha>0.85$), even when the number of questions retained changes dramatically \cite{gutzwiller2019positive,spain2008effect,Verberne2015,Beggiato2013}.  Issues have arisen between the negative and positive affect sides of the scale, leading some to posit that it consists of two sub-scales, one for trust and one for distrust \cite{jian,spain2008towards}.  On the other hand, the appearance of two sub-scales may be an artifact as negatively valenced items often cluster in factor analysis \cite{merritt2012two}.  

\begin{figure*}[]
    \centering
    \includegraphics[trim=1cm 1cm 1cm 1cm,width=\textwidth]{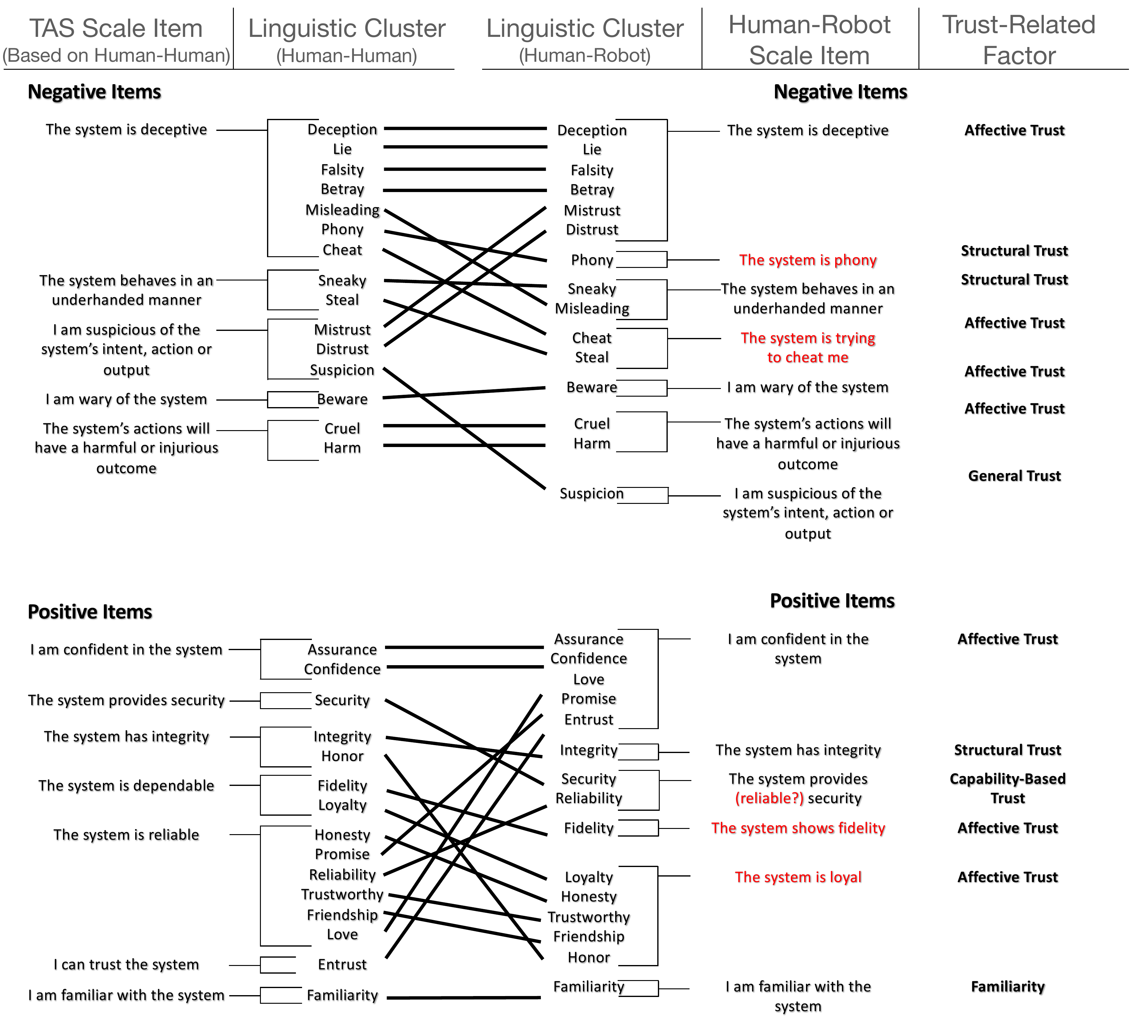}
    \caption{Re-Mapping Jian's Trust Items from the Human-Human linguistic trust clusters to the Human-Robot clusters per their own paper (Jian, 2000).  Our proposed items that do not appear on the original scale appear in red.  The bold lines indicate where terms mapped to when the trustee was assumed to be a machine instead of a human.}
    \label{fig:Jian_Groups}
\end{figure*}

Another significant issue is that the final version of the trust survey is based on the clusters found for their Human-Human Analysis rather than those identified in the General or Human-Machine Analyses.  Yet, it is the Human-Human-based survey that has become standard for assessing HRI and HMI \cite{Niu2018,Desai2012,Weitz2019}.  As can be seen in Fig. \ref{fig:Jian_Groups}, when comparing the result of the Human-Human to the Human-Machine analysis there, the clusters and the items they support, are not consistent.  Among the items of negative valence, two clusters are not inquired of at all in the current version of TAS (\emph{cheating}, \emph{phoniness}).  Furthermore, the shifting of `mistrust' and `distrust' from the \emph{suspicion} to \emph{deception} cluster may reflect their treatment as the inverses of \textbf{Affective Trust} instead of \textbf{General Trust}.  

Much more impactful are the differences in the items with positive valence.  The clusters behind the items' \emph{reliability}, \emph{dependability}, and \emph{trust} simply fall apart, completely losing any categorical identity with the other terms in their clusters.  Note \emph{dependability} is not even a term retained in the linguistic clustering of TAS but was reintroduced as the cluster name assigned to loyalty and fidelity.   Furthermore, in the human-machine clustering \emph{system integrity} loses its association with honor, suggesting that integrity has a different meaning in the machine context, perhaps relating more to soundness than a moral virtue.  Most importantly, at the item level, \emph{loyalty} replaces \emph{reliability}, and \emph{fidelity} becomes its own standalone item.  Finally, we must note the top-most positive valence human-robot cluster in Fig. \ref{fig:Jian_Groups} is very hard to label (here called \emph{confident}).  It seems to be a combination of \textbf{Affective} and \textbf{Structural Trust} - something closer to one's confidence in love-motivated promises or assurances.

The authors of \citeA{jian} posit 8 underlying factors of human-machine trust.  However, they admit to having difficulty distinguishing more than the first three \cite[p.21-23]{jian}: \textit{familiarity, reliability}, and \textit{confidence}.  Naming aside, these 8 factors explain 79\% of the variance in the survey.  The authors also note in the factor analysis that many positive and negative items load on the same factors, but not all of them, highlighting a partial divergence between trust and distrust constructs.  While cluster analysis initially revealed the two primary sub-scales, one of the positive and the other of the negative items, it also showed some finer resolution groupings depending on the correlational cut-off adopted.  Later studies in the field found that the 12-item survey supported the positive and negative sub-scales \cite{Hoffman2013,dolgov2017trust,holthausen2020situational}.  The finding that items group by effect is not surprising but does not speak much to the actual underlying constructs behind trust.

Overall, while TAS is currently the most popular scale, is easy to administer, and has strong internal and test-retest reliability, it could be improved by better construct validity, less terminological confusion, and scale items based on appropriate clusters.

\subsection{Instrument: Human-Computer Trust survey  (HCT or HCTS), (Madsen, 2000)}
\paragraph{Description.}
The Human-Computer Trust survey  (HCT or HCTS \cite{Madsen2000}) was the first validated technological trust survey widely utilized.  The HCT has five factors divided into two larger categories: \textit{affect-based trust}, composed of faith and personal attachment, and \textit{cognitive-based trust}, covering perceived understandability, technical competence, and reliability.  An essential contribution of HCT to trust theory was its surprising finding that \textit{affective-based trust} was stronger than \textit{cognitive-based trust}, though they did not quantify this result in their publication.  Even so, this finding challenged the traditional HRI trust research up to that time, which had almost exclusively focused on \textit{cognitive-based trust}.

\paragraph{Construct Mapping.}
We find several direct mappings comparing HCT's factors with the reference set.  Their \textit{personal attachment} maps nicely to \textbf{Emotional Response}, their \textit{understandability} to  \textbf{Shared Mental Model}, and their \textit{faith} to  \textbf{General Trust} (note a decidedly different usage of the term `faith' than in the reference factors).    

\paragraph{Validity \& Reliability.}
HCT, while highly cited, has only been tested for internal reliability, but not for less validity.  Madsen and Gregor \cite{Madsen2000} had proposed 5 factors, but factor analysis yielded 5 others that did not map cleanly to those initially proposed.  Even among those identified, the two weakest constructs, surprisingly, were their \textit{perceived reliability} and \textit{technical competence}.  These categories resulted in just two unique questions each, which is not enough to establish either as a validated factor.  One likely cause of their weakness was their sharing of quite a bit of language and sentence structure, which is known to cause a bias toward clustering together during factor analysis \cite{merritt2012two}. 

Beyond this weakness, the authors acknowledge other limitations, such as the low sample size of just 75, which was approximately a 3:1 ratio of subjects to survey items, and that only a single round of testing was performed.  Considering these issues, more validation was called for \cite{adams2003trust}.  A re-mapping to just the three stronger, validated factors with 88 participants in a different experiment was shown to have reliability, but validity analysis was still lacking \cite{Chancey2016TheEO}. 

Recently the HCT was re-tested for validity, as well as against Jian's TAS Scale \cite{dolgov2017trust}.  They confirmed that the HCT had high internal reliability, both overall and per factor ($\alpha>80\%$), and found high explained variance ($<70\%$).  However, the loadings indicated only four factors (not 5).  Unfortunately, these factors displayed low construct validity.  While this recent researcher did not do a full CFA, their work quantitatively confirmed some of HCT's main weaknesses. 

\subsection{Instrument: Gefen Instrument (Gefen, 2003)}
\paragraph{Description.}
The Gefen instrument \cite{Gefen2003} was developed by having experienced online shoppers perform a retrospective on the last e-vendor they encountered.  The authors settled on eight factors revised according to best CFA practices.  Their resulting instrument was responsible for introducing and popularizing many factor categories.  Being the first to combine the Technology Acceptance Model (TAM) with McKnight and Chervany's model of trust \cite{McKnight2001}, they can be credited for introducing  \textbf{Intended Use}, \textbf{Structural Assurance}, and \textbf{Situational Normality} to HAI/HRI.   

\paragraph{Construct Mapping.}
As this instrument was developed nearly independently from the HAI/HRI tradition, its terminology differs from previous works in the field.  A few critical nuances are worth special attention.  The closest Gefen comes to \textbf{Affective Trust} is what they termed \textit{Calculative-base Beliefs}.  However, these beliefs are best understood as the most cynical construct of \textbf{Affective Trust}, measuring a bare minimum of non-exploitation and non-malevolence, as opposed to active cooperation.  Gefen also introduced the concept of \textit{Perceived Ease of Use} and \textit{Perceived Usefulness} into the HAI/HRI trust measurement literature (e.g., \cite{Wang2005,Hegner2019}). \textit{Perceived Ease of Use} maps within the \textbf{Shared Mental Model} factor, as it is the main hypothesized result of having a \textbf{Shared Mental Model}, and is taken to indicate the strength of the human's confidence in the Shared Mental Model.  \textit{Perceived Usefulness} is more complicated; some have construed it to mean how useful something has been (\textbf{\emph{Learned Trust}}), whereas others have taken it to be anticipatory (\textbf{\emph{Situational Trust}}).  Depending on the instrument design, \textit{Perceived Usefulness} may alternatively be classified under \textbf{Faith in (Specific) Technology} or \textbf{Capability-based Trust} -- the latter being the case for the Gefen instrument.

\paragraph{Validity \& Reliability.}
This instrument achieved good factor reliability ($\alpha > 0.77$), and its creators not only performed tests of convergent and discriminate validity but also did a full structural equation modeling (SEM) to test for the plausibility of directional causation between sub-constructs.   The survey was first pre-tested with 50 participants, and the main validation was on a sample of 213 participants.  Interestingly, the survey was given as a retrospective and not focused on a specific technology.  Each participant answered regarding the last e-vendor with which they had interacted.

The instrument's main weakness is that clustered items exhibit highly overlapped phrasing within each factor, which may have biased their likelihood of clustering, internal reliability, and validity scores.  Additionally, after dropping items to achieve sufficient reliability, both the \textit{Familiarity} and \textit{Situational Normality} factors were left with only two remaining items, insufficient for establishing stable factor validity.  A number of the remaining factors displayed explained variances ranging from 0.44-0.62, where the suggested cut-off is generally $>0.6$, implying significant weaknesses in \textit{Perceived Usefulness} and \textit{Perceived Ease of Use}, as well as borderline acceptability for \textit{Trust and Intended Use}.  The authors note that a large proportion of their fit statistics were marginally below accepted standards, too.

Wang and Benbaset \cite{Wang2005} replicated a large part of Gefen's survey but for new users, as opposed to experienced ones, confirming good reliability of the factors but much weaker and lower explained variances and construct validity (as indicated by significant, high cross-loadings).  They also identified a different SEM structure, which provided a better fit.  Their structure, unlike Gefen's, implied that the various trust factors (\textbf{Affective}, \textbf{General}, \textbf{Capability-based}, etc.) are dependent upon the \textbf{Shared Mental Model}.  Differences in trust constructs between new and experienced users are expected, as trust takes time to initiate, develop, and calibrate \cite{Sollner2013}.  Still, it was surprising that not just the weights between factors but their actual inferred direction of influence changed between these proposed models.

A fuller critique of Gefen and TAM, on which it is based, can be found in the Discussion (Section \ref{Limits}).

 \subsection{Instrument: SHAPE ATM Trust Index (SATI; Goillou et al., 2003)}
  \paragraph{Description.}
SATI \cite{SATI} was developed specifically for assessing trust in air-traffic control systems and was founded on three pillars: \textit{trust}, \textit{situation awareness}, and \textit{teamwork}.  The trust aspect was heavily influenced by Madsen \& Gregor's HCT survey \cite{Madsen2000} and SATI, in turn, has influenced many additional surveys \cite{Chien2014,li2010cross,cummings2010supporting,Yagoda}. 


\paragraph{Construct Mapping.}
SATI is one of the most challenging scales to classify using our approach.  The main trust questionnaire contains two parts - the first is an expansion of Lee and Moray (\citeyear{Lee_94}) focusing on \textbf{General Trust} and asking about trust in each part of the system and team: self-confidence, trust in a specific technology, overall trust in the technology, as well as trust in colleagues, designers, and operators.  The second part consists of questions rating or ranking trust along 7 sub-scales and antecedents.  They do include \textit{Liking} (e.g., \textbf{Emotional Response}), \textbf{Familiarity}, and \textit{Understanding} (e.g., \textbf{Shared Mental Model}); with the rest of the items falling under \textbf{Capability-based Trust}, such as \textit{Reliability} and \textit{Accuracy}. A third instrument for assessing \textbf{Faith in Technology} is also proposed.

\paragraph{Validity \& Reliability.}
While SATI has no reported reliability or empirical validation, it does focus quite a bit on two other aspects of construct validity: face and item validity.  The authors formulated a list of 16 validation and success criteria, of which five were met, two were still being confirmed, and a further eight were being actively worked on.  As far as we could discern, a final report on achieving these criteria was never published.

Beyond validation and reliability, SATI  impacted the trust field in several important ways.  First, it highlighted the importance of ease of use for the instrument itself.  Second, it was an early proponent of the idea that not all sub-scales varied over time, preceding Hoff and Bashir's three-layered trust \cite{Hoff2015}.  Third, it revealed that system operators viewed trust as a binary decision and preferred to rate that decision independently of trust beliefs.  Furthermore, they found that confidence had to meet a certain threshold for the binary decision to trust to be made but that this threshold could be articulated by most.  Fourth, SATI unveiled that the type of failure mattered more than frequency.  Failures leading to accidents were perceived as far worse than false alarms, and a single accident was worse for trust than frequent false alarms \cite{SATI}.  Finally, it qualitatively demonstrated that monitoring and trust are not inverses, aligning it with other critiques of this assertion \cite{Razin2021b,castelfranchi2010trust} in Mayer's classic formulation \cite{Mayer1995}. 

 \subsection{Instrument:  Unified Theory of Acceptance and Use of Technology (UTUAT; Heerink, 2009)}
 \paragraph{Description.}
Heerink's Trust extension \cite{Heerink2009,Heerink2011} of the Unified Theory of Acceptance and Use of Technology (UTUAT) is similar to Gefen's extension of TAM\cite{Gefen2003}\footnote{UTUAT is itself built off TAM but started incorporating environmental, attitudinal, and \textbf{Emotional Response} factors much earlier}.  Heerink's survey posits 12 factors, with the primary goal of measuring the acceptance and use of a specific technology.  

\paragraph{Construct Mapping.}
UTUAT includes a factor called \textit{Social Presence} which aims to capture comfort/likeability, though its items are more along the line of \textbf{Situational Normality} than \textbf{Emotional Response}.  Its \textit{Facilitating Conditions}  maps directly to (perceived) \textbf{Familiarity}.  This was the rare survey that asked about \textit{Social Influence}, a factor identified with Reputation/Recommendation that in human-human trust is considered a sub-factor of familiarity \cite{yamagishi2001}.  We have classified \textit{Attitude} under \textbf{Capability-based} since it is very close to \textit{Perceived Usefulness}, but may also map to  \textbf{Faith in Technology}.  Finally, \textit{Perceived Adaptability} is clearly defined as goal-supportive helpfulness, and thus a form of \textbf{Affective Trust}.

\paragraph{Validity \& Reliability.}
UTUAT has undergone extensive reliability and validation analysis, although there are some remaining causes for concern, which will be discussed further in Section \ref{Limits}.  While very popular (see Table \ref{tab:overall_stats}), there are also concerns with the validity of Heerink's Trust extension of UTUAT.  Heerink's survey was only tested on two small populations (30 and 66), and only internal reliability data was reported.  The cut-off for Cronbach's $\alpha$ was kept low, at 0.7.  As no validation data is reported, it is unknown how distinct or meaningful many of the factors are.  Two of the factors, including \textit{trust} itself, are only composed of two questions.  A particular area of concern is that the methods used (correlational and regression analyses) to establish the structure between these factors created results that are not in line with any of the other regression or SEM results from other surveys (compare against \cite{McKnightD2011,Gefen2003}.  In particular,  the direction of plausible causality seemed reversed between General Trust and other factors, suggesting its results should be used with caution.


\subsection{Instrument: McKnight Trust Survey (McKnight, 2001)}
\paragraph{Description.}
A decade after their seminal multi-disciplinary review of the trust literature \cite{McKnight2001}, \citeA{McKnightD2011} created a trust survey instrument for specific technologies .  Through pre-tested card sorting and broad definitional reviews, they identified 12 factors (7 being trust-related), grouped into three main second-order conceptual categories. 

\paragraph{Construct Mapping.}
This survey covered every factor in \textbf{\textit{Learned Trust}}, including a whole range of sub-factor distinctions, 
e.g., \textbf{Faith in (General) Technology} which is about positive expectations of technology in general, and the \textit{Trusting Stance} which looks at lack of suspicion.  Other interesting sub-factor distinctions were \textit{Functionality} vs. \textit{Reliability} within \textbf{Capability-based Trust} and \textit{Intention to Explore} and \textit{Deep Structure Usage} within \textbf{Intention to Use}.  McKnight also proposed a second-order structure with three levels \textit{Propensity to Trust} (\textbf{\textit{Dispositional}}), \textit{Institution-Based Trust}, and  \textit{Trusting Beliefs in a Specific Technology} (\textbf{\textit{Learned}}).  

\paragraph{Validity \& Reliability.}
McKnight's reliability and validation analyses are some of the most thorough and yield some of the most robust results.  The survey was tested on a single large sample of 359 participants.  Internal reliabilities were all above 0.8, and the average explained variance ranged from 0.56-0.81, with the majority above 0.7.  The factor loadings per item revealed strong construct validity, with high primary loadings and no significant cross-loadings.  All the CFA and SEM statistics indicated a good fit.  They also found moderate to strong predictive validity for usage based on trust.  Further support comes from \citeA{Tussyadiah2020}, which used and re-validated McKnight's survey, finding almost identical correlations, regressions, and second-order structure.  They added the use of NARS \cite{nomura2008prediction} to the \textbf{\textit{Dispositional}} layer, showing it had a significant adverse effect on trust, as might be expected.

\subsection{Instrument: Merritt's Collective Set of Trust Scales (Merrit, 2008, 2011, 2015, 2019)}
\paragraph{Description.}
Merritt has been refining and creating multiple scales for measuring trust and its correlates over many years \cite{Merritt2008,Merritt2011,merritt2015measuring,merritt2019automation}.  Her group has produced six separate stand-alone scales such as \textit{Trust} (\textbf{Capability-based Trust}) \cite{Merritt2008}, \textit{Liking} (\textbf{Emotional Response}) \cite{Merritt2011}, \textit{Propensity to Trust} (\textbf{Faith in Technology}) \cite{Merritt2011}, the Perfect Automation Schema (\textbf{Dispositional}) \cite{merritt2015measuring},  All-or-None Thinking scale (\textbf{Dispositional}) \cite{merritt2015measuring}, and Automation-Induced Complacency (\textbf{Dispositional}) \cite{merritt2019automation}.  

\paragraph{Construct Mapping.}
Merritt's collective work has produced vital insights with respect to \textbf{Dispositional Trust} and its effect on \textbf{Capability-based Trust}.  High expectations did not significantly impact \textbf{Capability-based Trust} or its failure, but an all-or-nothing attitude severely affects trust repair.  Furthermore, \textbf{Faith in Technology's} ability to alleviate workload has a more substantial effect on \textbf{Capability-based Trust} than the need for monitoring, but both are significant.  Thus, Merritt's work has begun to tease out the sub-scales within \textbf{Faith in Technology} that provide a path forward but remain limited in their focus on \textbf{Capability-based Trust}.

\paragraph{Validity \& Reliability.}
 Each scale was developed in multiple rounds of medium to large sample sizes (69-500) and reported reliability and validity criteria as appropriate.  Merrit's six-item scale (Muir's+1 item) had internal reliabilities ranging from $\alpha \backsim0.87-0.92$.  They also performed a CFA and compared multiple SEMs looking at \textit{Propensity to Trust}, initial trust, and post-task trust as well as \textit{liking}, many of which displayed good fit statistics, though other validity criteria were not reported \cite{Merritt2008,Merritt2011}. Their \textit{Liking} and \textit{Propensity to Trust} scales demonstrated internal reliabilities of $\alpha\backsim0.8-0.89$ and $0.86$, respectively, and both fit well by their CFA and SEM.  The Perfect Automation Schema (\textbf{Dispositional}) had a weaker but still acceptable reliability score of $0.76$, and the All-or-None Thinking scale was at the very edge of acceptability ($\alpha=0.68$) \cite{merritt2015measuring}.  The Automation-Induced Complacency scale had two factors (\emph{Alleviating Workload} and \emph{Monitoring}) that were tested for internal reliability, split-half reliability, and with a confirmatory factor analysis. They achieved fairly good values across the board ($\alpha_{AW}=0.87, \alpha_{M}=0.79, \chi^2=84.33, df=34$, RMSEA = 0.08, CFI = 0.92, TLI = 0.89) though concluded that the two factors should be treated as separate scales and not two factors within a single construct. 

 \subsection{Instrument: Subjective Fluency Metric Scale, (Hoffman, 2013)}
 \paragraph{Description.}
The Subjective Fluency Metric Scale \cite{Hoffman2013} was designed to assess fluency in human-robot teams.  It was primarily based on the popular Working Alliance Inventory  (WAI) \cite{horvath1989development} that measures the patient-therapist `alliance' and is thus a human-human teamwork/support scale.

\paragraph{Construct Mapping.}  The WAI is comprised of a \textit{bond} subscale (\textbf{Shared Mental Model}) and a \textit{goal} sub-scale (\textbf{Affective Trust}). To the WAI, \citeA{Hoffman2013} added what they called factors of \textit{human-robot fluency} (\textbf{Capability-based Trust}), \textit{robot relative contribution}, \textit{trust} (\textbf{General Trust}), \textit{positive teammate traits} (\textbf{Emotional Response}), \textit{improvement} (\textbf{Capability-based Trust}), and some individual standalone items regarding commitment and cooperation (\textbf{Affective Trust}) \cite{Hoffman2013}.  

\paragraph{Validity \& Reliability}
Despite multiple uses and adaptations in the human-robot fluency community \cite{Hoffman2013}, validation work has yet to be reported, though strong internal reliability scores have been demonstrated \cite{Hoffman2013,Dragan2015}.  This scale is promising because it converges to our identified factors while arising from a sub-field that developed separately from the rest of the HAI trust literature and is heavily influenced by a completely different side of human-human trust. 

 \subsection{Instrument: Trust Perception Scale for HRI, (Schaefer 2013,2016)}
 \paragraph{Description.}
Leading one of the major groups working on conceptualizing human-robot trust and human-automation interaction for years, Parasuraman's and Hancock's groups developed and explored numerous ways to parse and categorize previous trust research, after numerous meta-reviews and analyses \cite{Parasuraman2000,Sheridan2005,Parasuraman2008,Hancock2011a}.  Their efforts, as far as trust measurement, were brought to fruition by Schaefer, who used their framework to develop her own trust model and survey instrument, the Trust Perception Scale for HRI (TPS-HRI)\cite{schaefer2013perception}, with further influence by Rotter \cite{Rotter_67}, Muir \cite{Muir1987}, TAS \cite{jian}, and SATI \cite{SATI}.  Schaefer cast a wide net as well, extensively citing not only the HRI literature but also the automation trust literature, providing one of the most valuable appendices of summaries for dozens of trust instruments.

\paragraph{Construct Mapping.}
The TPS-HRI found four factors, which were identified as \textit{performance-based functionality}, \textit{robot behaviors/communication}, \textit{task/mission-specific}, and \textit{robot features}. On careful review of each factor, we map \textit{performance-based functionality} to \textbf{Capability-based Trust} and \textit{robot behaviors/communication} as a combination of \textbf{Emotional Response}, \textbf{Shared Mental Model}, and \textbf{Structural Trust}.  We also suggest that what she termed \textit{task/mission-specific} should be understood as a failure of \textbf{Structural Trust}, whereas \textit{robot features} should be understood as dangers arising from failures of robustness \textbf{Capability-based Trust}.  It is evident, though, that these mappings are some of the least clear to our proposed factors.  However, as our factors generally match the other reliable and validated surveys, it is essential to clarify this mapping to understand how Schaefer's instrument and the oft-cited body of trust research behind it (\cite{Sheridan2005,Parasuraman2008,Hancock2011a}) fit into the rest of the field.

\paragraph{Validity \& Reliability.}
Schaefer produced a well-validated, multi-round study that checked and refined scales and tested for sub-constructs of trust.  She tested 630 participants in six rounds, though the scale itself was only validated with the final 102 participants in two rounds, reducing the set of 172 items to both 40-item and 14-item instruments. 

Regarding validation, she began by asking subject matter experts and then used the content validity ratio to identify item importance.  From here, she created both a 40-item and a 14-item scale and compared them against TAS \cite{jian}.  In her factor analysis, she used a Kaiser Criterion of eigenvalues $>1$ for truncation, an orthogonal varimax rotation, and a loading cut-off of 0.3.  The varimax rotation may have hurt her validation, forcing orthogonality on trust factors that are often understood to be heavily correlated.  We can also see this drop in her explained variance, which was 79.63\% overall but fell to 30.64\% after rotation.  The cut-off chosen for cross-loadings under 0.3 should be separately considered from the cut-off for primary factor loadings, which is often said to need to be above 0.7 \cite{McKnightD2011}.  Not a single item had a factor loading over 0.7; only 3 out of the 172 reached over 0.6.  Factor reliability and item communalities were not reported, and given their orthogonal rotation, there are no correlations between factors.  However, there were some cross-loadings, especially between items on the \textit{performance-base} and \textit{coordination/communication-based} factor.  

Schaefer's work was a significant effort that brought together a vast number of previous surveys into focus that had not received due attention. The large sample sizes and multi-round testing have lent this survey an air of validity that has led to significantly increasing use in the field.  Thus, it is critical to note that challenges remain with its reliability and validity to assess the factors that makeup trust.

 \subsection{Instrument: Cultural Trust Instrument (CTI), (Chien, 2014)}
 \paragraph{Description.}
The Cultural Trust Instrument (CTI) \cite{Chien2014} was primarily based on TAS \cite{jian}, HCT \cite{Madsen2000}, and SATI \cite{SATI}, with some influence from \citeA{McKnightD2011} among others that were more focused on \textbf{Faith in Technology}. 

\paragraph{Construct Mapping.}
The instrument was refined and tested in two rounds, the first of 45 participants and the second of 65.  The first reduced a set of 110 potential items to 59, and then a further 21 items.  Nineteen of these items mapped to four factors, three of which they termed \textit{performance expectancy} (\textbf{Capability-based Trust}), \textit{process transparency} (\textbf{Shared Mental Models}), and \textit{purpose influence} (\textbf{Faith in Technology}), keeping them closely aligned with the framework in Lee and Moray \cite{Lee2004}.  The fourth factor is left \textit{unnamed} and is solely composed of three negative valence items from TAS \cite{jian} that span \textbf{Affective Trust}, \textbf{Structural Trust}, and perhaps \textbf{General Trust}.  It is unclear whether the negative scale items are truly separate sub-constructs or cluster together solely because they are inversely scored \cite{merritt2012two}.

\paragraph{Validity \& Reliability.}
The EFA showed that these items fell into five internally reliable factors ($\alpha>0.7$), four of which showed good construct validity based on factor loadings from the EFA.  The final reliable but non-validated factor seems to be related to  `unreliability' though its correlation with the \textbf{Capability-based} factor is left unreported.  Their models also captured 50\% of the variance of general trust and 70\% of specific trust.  While less used overall, even as recently as this year, the CTI continues to be used in place of TAS, HCT, and SATI, with researchers citing its higher quality "rigor and empirical validation" \cite{Crawford2021}.  While the CTI is indeed somewhat better validated, this perhaps still does not warrant its use without further validation.

 \subsection{Instrument: Trust in Automation (German TiA) Scale (K\"{o}rber, 2018)}
 \paragraph{Description.}
The final scale worth consideration is the Trust in Automation (German TiA) scale  \cite{Korber2018}, whose popularity is impressive given its relatively recent date of publication.  

\paragraph{Construct Mapping.}
The German TiA followed \cite{Mayer1995,Lee2004} using three primary dimensions of trust, which they call \textit{Understandability/Predictability} (\textbf{Shared Mental Model}), \textit{Reliability/Competence} (\textbf{Capability-based Trust}), and \textit{Intention of Developers} \textbf{(Affective Trust}). They then try to map those three dimensions to Trust in Automation (\textbf{General Trust}).  They also test adding factors for \textbf{Familiarity} and \textit{Propensity to Trust} (\textbf{Faith in Technology}).

\paragraph{Validity \& Reliability}
They iterated their survey over two rounds, totaling 152 participants.  They found that despite the five factors proposed, only four could be validated with an EFA, which combined most of the items in \textit{Understanding/Predictability} with \textit{Intention of Developers}.  This scale had fairly complete validation: an excellent reliability analysis, and a well-done factor analysis, with an eye toward construct, surface, empirical, and predictive validation.  Unfortunately, \textbf{Familiarity} and \textit{Intention of Developers} were initially only composed of two items each, though after analysis \textit{Intention of Developers} was shown to factor with \textit{Understanding/Predictability}, resulted in only \textbf{Familiarity} being left with two items.  The fourth factor, \textit{Propensity to Trust}, had fairly low factor loadings, with all just below 0.6.

\section{Discussion}
\subsection{Picking an Appropriate Trust Instrument}

\begin{table*}[]
    \centering
    \resizebox{\columnwidth}{!}{%
    \begin{tabular}{l"c|c|c|c|c|c|c|c|c}
        Instrument & \# of items & \# of Factors &  Dispositional & Situational & Shared Mental Model & Learned & Intention &Reliability&Validity
        \\\thickhline
        HMI Trust Scale \cite{Muir1987}  &9&1&0&0&0&1&0&\cellcolor{yellow!40}Limited&\cellcolor{red!60}None\\\hline
        TAS \cite{jian}&12&1&0&1&0&4&0&\cellcolor{green!15}Good&\cellcolor{yellow!40}Limited\\\hline
        HCT \cite{Madsen2000}&25&5&0&1&1&2&0&\cellcolor{green!15}Good&\cellcolor{yellow!40}Limited\\\hline
        Gefen's TAM with Trust \cite{Gefen2003}&25&8&0&2&1&4&1&\cellcolor{green!15}Good&\cellcolor{blue!15}Decent\\\hline
        SATI \cite{SATI}&17&?&0&1&1&2&0&\cellcolor{red!60}None&\cellcolor{yellow!40}Limited\\\hline
        UTUAT with Trust \cite{Heerink2009,Heerink2011}&41&12&0&3&1&3&1&\cellcolor{green!15}Good&\cellcolor{yellow!40}Limited\\\hline
        Trust in Specific Technology \cite{McKnightD2011}&39&10&1&1&0&4&1&\cellcolor{green!15}Good&\cellcolor{green!15}Good\\\hline
        \begin{tabular}{@{}l@{}}Merritt's collected Trust-related Scales\\\cite{Merritt2008,Merritt2011,merritt2015measuring,merritt2019automation}\end{tabular}&28&4&1&1&0&2&0&\cellcolor{green!15}Good&\cellcolor{blue!15}Decent\\\hline
        Subjective Fluency Metric Scale \cite{Hoffman2013}&30&7&0&1&1&3&0&\cellcolor{green!15}Good&\cellcolor{red!60}None\\\hline
        TPS-HRI \cite{schaefer2013perception}&40 (14)&4&0&1$^*$&1$^*$&2$^*$&0&\cellcolor{green!15}Good&\cellcolor{yellow!40}Limited\\\hline
        CTI \cite{Chien2014}&19&4&1&0&1&4&0&\cellcolor{green!15}Good&\cellcolor{blue!15}Decent\\\hline
        German TiA \cite{Korber2018}&17 &5&1&1&1&3&0&\cellcolor{green!15}Good&\cellcolor{green!15}Good\\\hline
        \textbf{Maximum}&&&1&3&1&4&1&&
    \end{tabular}}
    \caption{Coverage of the core identified internal trust factors by the Top 12.  \# of items are those specific to trust and its factors. \# of Factors is the number the authors find or confirm, whether they match ours or not.  The next five columns indicate whether they have some number of factors in those layers, with the maximum number per column given in the final row.  Quality of empirical reliability and validity is rated in order of None, Limited, Decent, Good. }
    \label{tab:top12sum}
\end{table*}

When deciding how to pick an appropriate trust survey for an experiment or deciding whether to design your own, consider the following in Table \ref{tab:top12sum} and the following questions:

\begin{itemize}
    \item Are you interested in trust in general or also the factors that lead to trust?
    \item Are you administering the survey before, during, and/or after the experiment?  How interruptible/chunkable is the task?
    \item How much previous experience/exposure does the person have to the technology?
    \item Are you interested in trusting attitudes in general towards technology or a specific type of technology (dispositional) or interested in trust during an interaction or a specific relationship with one piece of technology?
    \item What is the expected effect size that trust will play?/What is the expected effect size on trust? 
\end{itemize}

Single-factor trust surveys \cite<e.g.,>{Muir1987,jian,Merritt2008} are appropriate for those interested just in trust as a single stand-alone construct as well as are generally shorter and faster to administer, making them well-suited for quick in-task trust probing.  If the effect size is expected to be large, fewer questions are needed \cite{Korber2018}, but more if it is expected to be small.  At least three questions are recommended if possible, though one along the lines of Merritt's Trust Scale \cite{Merritt2011} would be even better.  Furthermore, if trust in technology is not the opposite of distrust but is somewhat inverse to self-confidence, then separate questions on distrust, uncertainty, and self-confidence should all be measured in order to get a convergent estimate of internal trust.

Multi-factor trust surveys are appropriate both before and after a trust interaction, as opposed to during a task.  They tend to take longer, and the temporal dynamics of trust and the interaction should be accounted for when such instruments are chosen.  If one is interested in dispositional trust, then a number of well-validated Dispositional Trust-specific instruments are available.  The same goes for General Trust in a specific technology, e.g., trust in autonomous vehicles.  Even when one is only interested in general, dispositional trust, it is maybe worthwhile to probe affective, structural, and Capability-based factors, as these are all recursively fed back to one's General Trust attitudes. 

Differences arise if one is gauging trust in new users, experts, or trust over time.  As mentioned above, the upper-level factors such as \textbf{Faith in Technology}, \textbf{Familiarity}, and \textbf{Situational Normality} will matter more earlier on, especially at the stage of initial acceptance \cite{Merritt2008,Ekman2018}.  It may not be worth measuring these if the interval is greater than six weeks \cite{sollner2016longitudinal}, though over that period one could watch how these effects decrease as the effect of factors within Learned Trust increase.  Either way, multi-factorial instruments are recommended for such longitudinal studies \cite{Korber2018}.

A common question may arise as to whether Affective and Structural Trust are both worth measuring, as many may not see how they are relevant to their specific robot pick-and-place task in the lab, for example.  That the robot could be nefarious and have competitive goals (Affective Trust) or not follow rules (Structural Trust) is often not even considered).  Asking about deception can, in fact, elicit suspicion in both of these categories, an issue to which we will return.  Affective Trust is likely always at play, even if it is only in the narrowest calculative-based sense, i.e., \citeA{Gefen2003}.  The best way to conceive of Structural Trust is to go back to \citeA{luhmann}, the originator of the whole science of trust.  Structural, or what he called societal trust, is the abstraction of affective or personal trust.  Affective Trust is founded in the dyadic trust relationship and is rooted in personal knowledge, and is realized through individual attachment, bond, loyalty, and love.  Structural Trust universalized this in a world of strangers.  Instead of personal, special treatment customized to individuals, Structural Trust means to be fair through rules equally applied to all.  The direction of technological development has meant that robots and automation are usually seen as tools that have no knowledge of the individual, so they are usually bound by Structural Trust.  Newer systems that learn from individuals and are becoming more adaptive and personalized may be expected to be governed more effectively.  However, Structural Trust is often ignored in lab settings, whereas it has clear effects in the real world with regard to branding, attitudes toward tech companies, government and industry regulations, and insurance liability.  Affective Trust can be difficult to separate from structural at times since we buy, accept, and use technologies that we believe will help fulfill our goals, but we choose goals with enabling constraints that conform to societal norms.  Such questions are at the bleeding edge, at the border of science fiction, which has often dealt with villains who override ethical constraints in AI agents (violating Structural Trust) to align these AIs with their own personal goals (establishing Affective Trust).  In general, AIs and robots are assumed to be more fair and ethical than humans \cite{starke2021fairness}, demonstrating potential bias in Structural Trust.  Though others have shown that this is domain-specific; for instance, in a driving simulator, autonomous cars were presumed to follow the law as much as human drivers \cite{razin2019}.

As noted above, even asking about or mentioning trust can plant seeds of suspicion.  This is the paradox of trust signaling.  Explicit displays of trust can build the trustee's confidence in the trusting relationship, whereas explicit mentions of trustworthiness can poison the well and lead to uncertainty and potential distrust.  This is a potential bias of all explicit trust survey instruments.

\subsection{Choosing and using an Appropriate Instrument}
What we have been working toward throughout this work is an attempt to characterize the current state-of-the-art so that this work can ultimately help guide practitioners in choosing an appropriate instrument and even designing future ones.    An excellent start to this discussion has already been provided by \citeA{Korber2018} and we do not intend to reproduce it here in full.    Instead, we aim to provide a simple guide for quick and easy reference.

\begin{enumerate}
    \item Single Factor: If you are interested in capturing trust as a single factor during an interaction, whether its antecedents simply do not matter to your application OR if you need to capture trust repeatedly over an interaction and therefore need to minimize interruption time, then a short survey is appropriate. 
    
    \begin{enumerate}
    \item Choosing a Pre-Existing Survey: Given the validity issues of Jian and Muir, Merrit's trust scale is a reasonable choice, though it is entirely focused on \textbf{General} and \textbf{Capability-based} trust and not suitable for more socially situated or effective applications.    Their short \emph{Liking} scale can be used to capture \textbf{Emotional Response} and their various short \textbf{\emph{Dispositional}} scales can be used for \textbf{Faith in Technology}. However, even Merritt's scales need further reliability and validation testing, and it would be useful for future works to carry these out and properly report them.
    
    \item For Designing and Assessing New Surveys: Any short one-factor survey must ensure to include at least three items
    \begin{enumerate}
        \item Construction: Each item must use different enough wording
        \item Reliability: Minimally Cronbach's $\alpha$ and McDonald's $\omega$ and the inter-item correlation, if not other methods such as test-retest
        \item Validity: Factor analysis is performed, and the items' loadings are sufficient on that single factor, generally $> 0.45$, though at least 4 loadings over 0.6 is preferred (if 3 items all, 3 should be over 0.6)
    \end{enumerate}
     
      Other items to be reported: 
      \begin{enumerate}
          \item Correlation matrix of items
          \item A collinearity check using the variance inflation factor ($< 5$)
          \item Kaiser-Meyer-Olkin measure of sampling adequacy ($> 0.6$)
          \item Loadings and communalities for each item on the single factor
          \item The explained variance
      \end{enumerate}
        A more complete list of recommendations, including sample size, treatment of missing data, variability, linearity, normality, and other methodological considerations, can be found in \citeA{watkins2018}.
    \end{enumerate}
    \item Multiple Factors: Sometimes, a more in-depth understanding of the {\fontfamily{qzc}\selectfont internal}, psychological factors of trust, and its antecedents are desired.  This sort of survey takes more time and is best suited for prior- and post-interaction administration as they tend to range from 15-45 questions.    This survey type is recommended for research questions such as how an interface design affects \textbf{Situational Normality} or an \textbf{Emotional Response} and therefore trust, how \textbf{Faith in Technology} influences final \textbf{Intent to Use}, or how programming law-following into an autonomous car influences \textbf{Structural} and \textbf{ General Trust}.
    
    \begin{enumerate}
    \item Choosing a Pre-Existing Surveys: At the time of writing, McKnight's survey is the strongest and most complete of the Top 12 in that it covers the most factors with the best validity.  A shorter and more limited choice would be K\"{o}rber's survey (including their additional survey for \textbf{Faith in Technology}). Sometimes though shorter can be preferable.  K\"{o}rber's is mainly limited by its lack of \textbf{Structural Trust}.  From our general survey of the 62 instruments, a few of the most recent instruments are very promising and provide broad coverage of the identified factors, though lacking verification of reliability through repeated use \cite{Park2020,Chi2021}.
     \item For Designing and Assessing New Surveys: Almost all of the same advice from the one-factor survey development applies here, except the number of factors retained must be justified and multicollinearity is instead indicated by cross-loadings $>0.3$.    In addition, Bartlett's Test for Sphericity should be significant and reported, as should the rotation used, item uniqueness, and the explained variance of each factor. 
     
     If a confirmatory factor analysis or structural equation model is performed, then several additional elements should be carried out and reported \cite{jackson2009reporting}:
     \begin{enumerate}
         \item Specification of multiple prior models to test
         \item Theoretical justification of those various models
         \item Choice of estimation procedure.    If the method used assumes multivariate normality, such as Maximum Likelihood (ML), then normality reporting is even more critical.
         \item Reporting of fit tests for each model: $\chi^2$ and root mean squared error of approximation (RMSEA) are most important, but additional indices such as the Comparative Fix Index (CFI), Tucker-Lewis Index (TLI), and Standardized Root Mean Residual (SRMR) are also highly recommended.    CFI and TLI should minimally be $>0.9$ though more recent recommendations range from 0.95-0.97.    However, strict cutoffs are also advised against.    RMSEA and SRMR should at least be below 0.1, though under 0.06 is preferred by many.
         \item Other elements to report: Parameter estimates, the variance of exogenous variables with standard errors, explained variance by endogenous variables, structure coefficients
         \item Finally, the preferred model and justification of it both on the grounds of its quality of fit as well as theoretically.
     \end{enumerate}
     A full checklist for CFA reporting can be found in the Appendix of \citeA{jackson2009reporting}.    If an SEM is reported, researchers must understand to what extent an SEM supports claims of causality and what assumptions such an analysis entails \cite{bollen2013eight}.
    \end{enumerate}
\end{enumerate}

\subsection{Outstanding Issues}


\subsubsection{External Trust Factors}
While the extant surveys converge on the trust model presented above, there is good reason to believe from numerous experiments that other factors may have some impact.  While the literature has explored dozens of potential factors \cite{Hancock2011a,Hoff2015}, only a few have thus far withstood some rigor and repeatability, as shown in Fig. \ref{fig:expanded_trust}.  We will not attempt a complete review of these additional factors here.  However, we will outline their potential roles and give an overview of how they interact with the emergent validated trust model.

Continuing to follow the three-layer schema of \citeA{Hoff2015}, these additional factors may be divided into three main categories: expanded dispositional trust factors or antecedents, external situational trust factors, and external learned trust factors.  We introduce the language of externality here to indicate that these factors are not part of the internal trust model and are directly contingent on the environment and embodiment of the agents in that environment.  This is clearest for robot form, which is a physical property of the actual technology.  Monitoring is also external in the sense that it is the direct sensory link between the human, the technology, and the environment.  Perceived risk is likewise an emergent property of this three-way interaction.  Cognitive load is more indirect but serves as a constraint on cognitive processing, and while it affects trust is external to it.

On the other hand, external dispositional antecedents of trust are not directly trust-related and perhaps are best understood as the foundations of trust that arise out of those factors outside of the individual's control, such as their age, culture, personality, which in turn give rise to their group membership, education.  These shape trust-related factors directly more directly, such as the willingness to take risks and one's faith in people, both in individual strangers as well as institutions.  These expanded dispositional factors are discussed in detail in \citeA{Hoff2015}, with the role of Personality getting a more complete and up-to-date treatment in \citeA{alarcon2021role}.  Short, updated treatments of the remaining areas are presented below.

\subsubsection{Gender}
Multiple studies to date have not found gender to have a significant effect on trust in automation or robots \cite{schaefer2013perception} while others have shown mixed effects.  For instance, in one study where men had more experience with computers, they unsurprisingly found them easier to use but not necessarily trust \cite{Heerink2011}.  On the other hand, another study found that women trusted a robot more, perhaps because men had a higher need for cognition \cite{Robert2009}, thus indicating a mediating effect on the relationship between confidence in the robot and self-confidence.  Adding further complexity, \citeA{kuchenbrandt2014keep} found that gender may have a mediating role in setting group expectations.  One study indicated that a weak interaction effect occurs for men interacting with a gendered robot but not women.  In all cases, when gender has been found to have any sort of effect, it has been weak.  It is also unclear whether gender has an immediate effect on trust or just mediates training, group membership, and cognition.

\subsubsection{Age}
Age has shown a clearer effect than gender on trust.  Aging, for instance, decreases willingness to take risks \cite{dohmen2018identifying}, which in turn decreases willingness to trust \cite{Desai2012}.  On the other hand, age lends familiarity \cite{sundar2016hollywood}, and more experience trusting in general.  This, in turn, translates into better trust calibration and an understanding of what factors in the interaction actually matter \cite{Hoff2015}.

\subsubsection{Culture}
The role of culture in trust of technologies has also seen limited exploration thus far.  Culture, when it is studied, has been taken in the most narrow, reductionist sense - either as nationality \cite{Chien2014,yerdon2017investigating} or along the axes of horizontal vs. vertical achievement and individualist vs. collectivist \cite{huang2017users}.  Similar to the findings on gender, culture shapes perceptions of group membership \cite{wagner2015robots}, faith in others \cite{yamagishi2001}, and exposure/education, and thus familiarity.  The same cultural dimensions on which men score higher are those that lower trust, indicating a complex interaction of culture and gender \cite{zhang2011effect,zeffane2020gender}.   Cultures also shape general attitudes toward technologies \cite{Chien2014}, but through access as well as portrayal in the media, as illustrated by the Hollywood Robot Syndrome \cite{sundar2016hollywood}.  When culture has been shown to affect General Trust, it has been weak and further weakens over time, like other dispositional factors \cite{Chien2014}.

\subsubsection{Faith in People}
Like Faith in Technology, Faith in People seems to have a weak effect on Intention to Use, and its effect on General Trust lessens with the usage of the specific technology \cite{uslaner2015measuring}.  However, many of the experimental designs have essentially excluded the more salient aspects of Faith in People by not looking at it through the lens of dispositional institutional trust, which serves as the foundation of specific Structural Trust.  (This being a primary focus of human-human trust surveys deriving from \citeA{Rotter_67}) Faith in people may also set the \textit{a priori} standards expected of Capability-based Trust, shared mental models, and familiarity, as it creates a baseline against which to compare technology and confidence in the designers to be ethical, follow known standards, and design for normative interactions \cite{razin2019}.

\subsubsection{Group Membership}
At the root of trust from the cognitive perspective lies group membership \cite{williams2001whom}.  The famous trust `sniff test' found that intranasal oxytocin increased in-group trust but decreased out-group trust \cite{van2012sniff}, indicating that propensity to trust is deeply entwined with the moral dimension of In-Group Loyalty \cite{haidt2007new}.  Furthermore, when this experiment was tried in an HRI context, they found that oxytocin specifically increased trust in the `uncanny valley' \cite{de2017little}, which fits well with the finding in human-human trust that the Big 5 personality trait of Agreeableness has a strong effect on trusting strangers \cite{freitag2016personality}.  Furthermore, group membership serves to form prior expectations of Familiarity, Situation Normality, and Emotional Response that proceed to form the shared mental model necessary to capture the relevant qualities of the trustee \cite{tanis2005social}.  Such stereotyping has been exploited in HRI trust experiments by \citeA{wagner2015robots}, who uses it to literally extract reward, goal, and capability beliefs in forming prior trust expectations.  Furthermore, expectations associated with in-group hierarchies regarding age and gender seem to play an important role in trust biasing \cite{pak2012decision} and have a direct impact on robot form.

\subsubsection{(Robot) Form}
Trust's interaction with robot form is a popular area of study, with much work focused on anthropomorphism.  However, results concerning the effect of form on trust have had some mixed results \cite{hauslschmid2017supportingtrust,waytz2014mind,de2017little}.  It seems that at least in the cultures thus far studied, there is a general initial overtrust of automation above and beyond human-human overtrust.  This overtrust stems from a belief that machines are less biased, fairer, and thus more just \cite{starke2021fairness}.  When this belief is uprooted, trust repair is much harder.  In part, this is explained by the `higher they are, the farther they fall' in terms of miscalibration.  Making the robot more human-like can increase trust resilience under uncertainty and better calibrates initial over-trust \cite{de2017little}.   However, other results have shown what seems to be an `uncanny valley' of trust, where seeming too human also lowers perceived trustworthiness \cite{hauslschmid2017supportingtrust}.   These results are further complexified by the multitude of interactions mentioned above between age, culture, gender, personality, group membership, and educational background that shape how forms are perceived and interacted with.  
In terms of the emergent validated mode, anthropomorphism is thought to invoke Situation Normality and Emotional Response that allow for the creation of Shared Mental Models by helping people determine the generalized model by which to start forming expectations. 

Anthropomorphism is not the end of the story, though.  Form shapes the user interface by which the human can assess and monitor the trustee as needed.  We simply refer the curious reader to the vast literature which already exists on interface design for trust \cite{wang2005trust,wang2005overview,Skarlatidou2013,Wojton2020}, and legibility \cite{Dragan2015}.

\subsubsection{External Learned Trust Factors}
The three primary factors that seem to affect learned trust are \textit{cognitive load, perceived risk, and monitoring} which supports the core of the feedback loop.  The ability to monitor the perception of data \cite{morra2019building}, behavior \cite{xu2015optimo},  or via explainable AI \cite{adams2005human} are crucial in supporting the full range of learning within learned trust.   Monitoring of the system both adds to cognitive load and can affect learned trust in all sorts of ways, updating estimates of goals, capabilities, loyalties, and rule-following as well.  One of the best-studied external factors that influences learned trust is cognitive load.  Almost always assessed by the NASA TLX \cite{hart1988development}, cognitive load has been shown to increase intent to use \cite{Desai2012}.  It is important to note that workload does not significantly change due to the system's capability/accuracy \cite{Dadashi2013, deVisser, Wang2018}, but by the quality of the information provided as feedback during monitoring \cite{Dadashi2013,helldin2014transparency}.  As tasks get harder to monitor, frustration and effort increase \cite{helldin2014transparency}, self-confidence decreases, and the need to trust in the technology is essentially coerced into complacency \cite{IdramaniL.SinghRobertMalloy1993}.     In some ways, perceived risk has a similar effect on intent to use as cognitive load if the risk makes the task more difficult \cite{Yagoda}.  However, as the potential goals of the trust themselves have more extreme payoffs/costs associated with them, the more need there is for some combination of monitoring, self-confidence, and trust in the system to compensate.  One critical area that needs more attention is the interaction between willingness to risk, perceived risk, and types of risk.

The external antecedents and factors that influence trust are by no means exhaustive.  However, we feel that the ones described above present the next frontier for validation and testing beyond the `internal' emergent validated trust model.  The model we have presented, therefore, is meant to serve as a foundation for integrating past research into the converging language of trust in technology and automation while directing future research to attend to the myriad interplays of external factors with trust.

\subsubsection{Missing Internal Trust Factors}
While we have attempted to create broad categories to describe the results of the emergent validated trust model, there are nuances and differences within factors that we have not fully explored, which are left for future work.  

As we have already discussed, there does seem to be a general and detectable distinction between expectations and confidence in those expectations; this is best seen by the dichotomy in many \textit{Capability-based Trust }measures, where competence/performance/capability are often found to be distinct from reliability/predictability.  While less clear, a similar distinction may be evident within \textit{Structural Trust}, between the sub-factors of \textit{ethical} and \textit{sincere} \cite{malle2021multidimensional}.    There are also some important distinctions expected within \textit{Structural Trust} as to whether it is internally or externally motivated \cite{dunning2014trust}. 

It is still somewhat unclear whether the distinction between \textit{Propensity to Trust} and \textit{Trusting Stance} is fully warranted. 

The \textit{Situational trust} factors have also not been tested together for validity and reliability, and their correlational, and potentially causal, structure is still very unclear.  It may be that \textbf{Familiarity} and \textbf{Situation Normality} are sub-sets of one another.  It is interesting to note how much they co-vary with \textbf{Emotional Response}, speaking to our intuitive understanding that how much comfort or liking we 
have is dependent on how familiar or normal an agent or interaction appears to be.

\textbf{Emotional Response} may be too broad as a single category, but it at least seems that the factors of emotional attachment, liking, enjoyment, warmth, and engagement may be distinct but overlapping in the roles they play in trust development.

Perceived Usefulness has also proven to be a complex category, but mostly due to its textual ambiguity.  Many take it descriptively\textemdash how useful something is\textemdash  in the vein of \textbf{Capability-based Trust}.  However, some authors have used the term prescriptively, as in how useful something has the potential to be.  One can see from the extended general model that Perceived Risk is accounted for but not potential gains or reward.  The need to account for value-added benefit has been specifically called out in UTUAT critiques \cite{Shachak2019} but has received little to no attention in the robotics or automation world in this explicit survey domain.  Alternative game theory-based approaches using interdependence theory have begun to shed light on the role of relative expected gains \cite{Razin2021b}.

\subsubsection{Distrust}
This model does not explicitly deal with distrust.  The debate over the nature of distrust and its relationship with trust is less than clear.  Ever since \citeA{Lewicki1998} posited that trust and distrust were not opposites and should not be measured on the same scale, their relationship with one another has become blurred. 

\citeA{Lee2004} essentially divorced distrust from the trust calibration conversation, as it was neither over nor under trust.  The opposite of trust can better be categorized as a combination of distrust and uncertainty.  But distrust in the trust literature is often confined very narrowly in the sense of negative Affective Trust (malevolence) or suspicion of such (the negative valence confidence sub-factor of Affective Trust).   \citeA{mcknight2001while}, and \citeA{dimoka2010does} have given strong descriptions of models that test distrust from a triadic perspective (structural, capability, and affective).  An interesting inversion was attempted by \citeA{mcknight2006distrust}, where they tested a positively framed trust survey against an equivalent negatively framed one.  While \textit{insufficient reliability} and \textit{insufficient validity} measures were published on the `Negative' survey, they demonstrated that the negatively valenced survey actually may be more sensitive and have higher explained variance.  This is less a question of the underlying model as much as the framing of said model in its measurement.  It also speaks to the imbalance in the trust signaling problem.

\subsection{Comparison with Human-Human Trust Measurement}
While early works on human-human trust were entirely focused on dispositional trust measures \cite{Rotter_67, gillespie2003measuring, mcevily2011measuring}, most focused on political/societal trust more generally.  Measures for inter-personal trust started developing in many professional contexts, from organizational trust between employees and managers, between firms, and within business networks \cite{Mayer1995,bhattacherjee2002individual,schoorman2007integrative,johnson2005cognitive}.  A second major approach developed, which studied trust in friendships and intimate relationships \cite{gottman2011science,bukowski1994measuring}.  Building off these approaches, other veins developed in the healthcare community between patients and doctors \cite{thom2004measuring,anderson1990development}, as well as trust in media \cite{matthes2008content}, and trust in strangers \cite{ermisch2009measuring}.  

Those studying HAI, HCI, and HRI trust have often turned to human-human studies both directly and indirectly.  Fluency scales \cite{Hoffman2013} are primarily built off the Working Alliance Inventory originally developed for patient-therapist trust \cite{horvath1989development}.  Other HRI studies simply lifted scales from a doctor-patient trust (\citeA<e.g.,>{Mann2015} using \citeA{anderson1990development}).  Many still cite or even test with \citeA{Rotter_67}, and the influence of \citeA{Mayer1995}, who primarily worked in organizational trust, is hard to overstate.

All branches of human trust measurement are plagued by similar issues as HAI and HRI trust.  A similar review to our own here was performed on human-human trust scales \cite{mcevily2011measuring}.  Interestingly, they identified approximately the same number of trust scales in human-human trust, and a very similar pattern, where 60\% had created their own ad hoc measures and 40\% re-used a previously validated instrument.  They similarly discussed the reliability and construct validity patterns over time.  During their review, they found even fewer of the human-human scales had reported empirical validity measures.  They found only 22\% had considered multi-factorial trust as opposed to single uni-dimensional items.

Of the dimensions reported, they fall closely in line with our own, as can be seen in Table \ref{tab:hu-hum}. 


Beyond a general similarity in factors, the similarity in structures has become more than apparent.   There was only one major grouping that our model did not capture, which was how available, open, and receptive the potential trustees and trustors are.  Likely, this grouping is closely tied to the importance of Agreeableness in trust from the Personality research \cite{freitag2016personality} as well as our Emotional Response factor.  However, it seems distinct enough within human-human trust to merit its own factor and a potential focus of future work.

Human-human trust is seeing similar re-analysis as to the importance of the shared mental model and mutual modeling in teams.  It introduced the great divide between affective- and Capability-based Trust (the latter they often call cognitive), as well as the trust-trustworthiness calibration problem.  Similar attempts at consensus and convergence are apart across both fields.  While many have considered the deep differences between human-human and human-robot trust, the work of measuring the factors of trust is extremely similar, to the point that we plan more general works to help bridge these gaps in the future.

\subsection{Limitations}
\label{Limits}
Gefen's work was a Trust extension for TAM, which has seen its own extensive testing and validation.  A complete review of the history, development, meta-analyses, strengths, and weaknesses of TAM can be found in \citeA{Chuttur}.  That work brings up many critical limitations  - that TAM does not account for relationship dynamics, that the connection between intention and behavior is not direct, that dispositional and Emotional Response factors are lacking, and that voluntary vs. mandatory use leads to differential strengths of Capability-based and Structural Trust.

Similar critiques of UTUAT exist to those of TAM \cite{Ammenwerth2019, Shachak2019}.  Both can explain actual use only so well (explained variances for intended use range from 0.3-0.7), and their power of explainability seems to have plateaued \cite{Ammenwerth2019}.  Neither is meant for measuring acceptance of mandated technologies, only voluntary ones.  Both have seen a good deal of instability in the language of their survey instruments.  Both are meant to be used as one-off assessments, usually post-task, and do not measure temporal changes \cite{Chuttur, Shachak2019}.  Several questions surround their assumptions.  Furthermore, Perceived Usefulness has proven more consistent in predicting Intended Use than Perceived Ease of Use or Social Influence, leading some to question their validity \cite{Ammenwerth2019}. 

Many of these same concerns have been voiced concerning survey instruments \cite{Korber2018}.  Wide-spread issues with validity, unstable language, cherry-picking of items to form new factors, and lack of attention to measuring temporality are not confined to TAM and UTUAT.  We have endeavored to show how Intent to Use is influenced by many factors beyond trust, and do not see this as a fundamental critique of the emergent validated model as such, whose purpose is to show the convergent model of trust, as opposed to these models which attempt to fully capture acceptance and usage.

\subsection{Trust as attitude, beliefs, intentions, decision, and behaviors}
This work has solely focused on capturing trust attitudes, beliefs, and intentions.  Many have pointed out the divide between trust beliefs and actual decisions and behaviors.  This is partly addressed by the external learned factors that affect trust.  Trust alone has proven insufficient to predict use \cite{dunning2014trust,Gefen2003,Desai2012}, as risk, need, coercion, and stress all interact with trust in complex ways.  Elsewhere, we have presented an alternative approach to trust decision-making, drawing from game theory, which is not covered herein.   There we present successful models that both estimate and predict trust in discrete situations using payoff calculations that we have argued reflect a trust index, as well as commitment, coercion, and coordination metrics.  In those works, we have argued that these indices and metrics interact with calculable thresholds for trust decision-making.  While much work remains to be done, clear lines can be drawn between the factors of trust we have outlined above and these measures.  Capability-based Trust allows for setting the payoffs, and the shared mental model and familiarity allow us to form probabilistic expectations of fulfillment.  Affective and Structural Trust both shape perceived commitment to goals as well as coercion through norms, requirements, and sanctions as well as incentives.  Dispositional trust allows us to form prior beliefs from which to refine our estimates, and willingness to risk supplies a threshold.  Cognitive workload can increase our commitment by decreasing our alternative options.  More work is needed to capture how these interactions play out and especially in how to measure perceived risk and reward, but progress is being made in explicating trust beliefs and trusting actions.  A major step forward would be a reliable, valid, multi-factorial trust survey instrument that can be used to measure trust, both broadly as well as temporally, as it evolves over time (forthcoming).

\section{Future Work and Conclusion}
This work has described the current state of measuring trust between humans and various technology categories.  It has assessed the types of metrics used, especially in terms of reliability, validation, and factor terminology.  The latter allowed us to identify ten major trust factors, and the accompanying meta-analysis helped us discern the general structure.  We also discussed how trust should be measured going forward, what areas of the model need further examination, and potential avenues to explore - from external trust factors and internal nuances to how trust beliefs become decisions.  By evaluating and integrating the results of these previous survey instruments and constructs, the model presented in this work provides a robust foundation for the future of human-machine trust research.

\section*{Acknowledgments}
We want to thank Matthew Brzowski and Dr. Dan Nathan-Roberts of San José State University for providing us with their raw data set.

\section*{Key Points}
\begin{itemize}
 \item  Both modeling and measuring trust have proven difficult due to construct proliferation, terminological mismatches, and the tendency of researchers to create their own survey instruments from scratch
 \item While many trust questionnaires lack reliability and validity, we can use the ones that are reliable and valid to extract reliable and valid factors.  Furthermore, the reliable and validated factors can be mapped across surveys based on the similarity of items within the factor, clearing away the terminological confusion
 \item Once factors are mapped, a consensus can be seen around the structure of trust, and how the factors affect each other, resolving much of the construct proliferation
 \item The most popular and impactful trust questionnaires are assessed for reliability and validity and compared against the commonly-shared factors and consensus model.  In general, this emergent model provides guidance for future development and testing of the model, as well as choice and creation of future survey instruments
\end{itemize}

\bibliographystyle{apacite}
\bibliography{biblio2}

\section*{}
Yosef S. Razin is a Ph.D. candidate in Robotics at the Georgia  Institute of Technology's School of  Aerospace  Engineering.  He earned a BSE in Mechanical and Aerospace Engineering from Princeton University in 2011.  The focus of his work is on human-robot interaction and human-automation trust.\\

Karen  M.  Feigh is a professor at the Georgia Institute of Technology's School of Aerospace Engineering.  She earned a Ph.D. in Industrial and Systems Engineering from the Georgia  Institute of Technology.  She leads the Cognitive Engineering Center, focusing on decision support and incorporating computational cognitive modeling in engineering design.

\newpage
\singlespacing
\onecolumn
\appendix
\setcounter{table}{0}
\renewcommand{\thetable}{A\arabic{table}}
\vspace{3in}
\begin{center}
    {\Huge Appendix A}
\end{center}

\begin{landscape}

\begin{longtable}[c]{l"c|c|c|c|c}
    \caption{Mapping Terms between Identified Dispositional and Situational Trust Factors of Significant Reliability and Validity.  Bolded authors are among those in the Top 12 survey instruments.  Entries in red were found to be reliable and valid but span more than one factor.}\\
     \label{tab:termMap1}
   \textbf{Author} &	\textbf{Faith in Technology} &	\textbf{Emotional Response} &	\textbf{Shared Mental Model} &	\textbf{Situation Normality} &	\textbf{Familiarity} \\\thickhline
CPRS: \citeA{IdramaniL.SinghRobertMalloy1993} &	Propensity to Trust	&&&&			\\\hline
\textbf{\citeA{Madsen2000}}	&&	Personal Attachment & \begin{tabular}{@{}c@{}}	Perceived Understandability,\\Technical Competence\end{tabular}&	&\\\hline
\textbf{\citeA{Gefen2003}}	&&&	Perceived Ease of Use&	Situation Normality&	Knowledge-based Familiarity	\\\hline
\textbf{\citeA{Korber2018}}	&Propensity to Trust&&		\begin{tabular}{@{}c@{}}Understanding/Predictability,\\Intention of Developers\end{tabular} &	&	Familiarity	\\\hline
\citeA{Scopelliti2005}&	\begin{tabular}{@{}c@{}}Attitude Towards\\ New Technologies\end{tabular}&	\begin{tabular}{@{}c@{}}Emotional Response\\ to Robots\end{tabular}&	\begin{tabular}{@{}c@{}}\color{red}Human-Robot\\ \color{red}Interaction\end{tabular} &\begin{tabular}{@{}c@{}}\color{red}Human-Robot\\ \color{red}Interaction\end{tabular}&	Capabilities of Robots	\\\hline
\begin{tabular}{@{}l@{}}\citeA{Wang2005}\\\citeA{Komiak2006}\end{tabular}	&&Emotional Trust&		Perceived Ease of Use	&&Familiarity\\\hline
\textbf{Heerink (2009, 2011)}& & \begin{tabular}{@{}c@{}}Anxiety,\\ Perceived Enjoyment,\\Perceived Sociability\end{tabular}&	Perceived Ease of Use&Social Presence&\begin{tabular}{@{}c@{}}Facilitating Conditions,\\Social Influence\end{tabular}\\\hline	
\textbf{\citeA{McKnightD2011}}&	\begin{tabular}{@{}c@{}}Faith in General Technology,\\Trusting Stance\end{tabular}&&	&	Situation Normality&\\\hline
\textbf{Merritt et al. (2008, 2011a)}	&Propensity to Trust &	Liking Scale&&&\\\hline
\citeA{Sollner2012}	&&	&	Process (formative)	&&	\\\hline
\textbf{\citeA{schaefer2013perception}}&&\begin{tabular}{@{}c@{}}\color{red}Robot Behaviors +\\\color{red}Communication\end{tabular}	&\begin{tabular}{@{}c@{}}\color{red}Robot Behaviors +\\\color{red}Communication\end{tabular}&&			\\\hline		
\citeA{Sollner2013}	&&&&&	\\\hline
Wechsung et al. (2013)&&		\begin{tabular}{@{}c@{}}Likeability,\\Entertainment\end{tabular}&&		Naturalness		\\\hline
Ullman et al. (2014, 2021)&&		Intelligence, Other&	&&			\\\hline		
\citeA{cherif2014impact} 	&&	Social Presence	&&&		\\\hline
\textbf{\citeA{Chien2014}}	&Purpose Influence&&		Process Transparency	&&			\\\hline
\citeA{Lee2015}	&&	Perceived Warmth&&&	\\\hline					
\citeA{Tussyadiah2020}&	\begin{tabular}{@{}c@{}}Faith in General Technology,\\Trusting Stance, NARS\end{tabular}	&&&&					\\\hline
\textbf{\citeA{Hoffman2013}}	&&	\begin{tabular}{@{}c@{}}Positive Teammate\\ Traits\end{tabular}&\begin{tabular}{@{}c@{}}Working Alliance-\\ Bond Subscale\end{tabular}&&	\\\hline
\citeA{Rupp2016}&	&&		Wearable Technology Trust	&&\\\hline
\citeA{Hegner2019}&	Personal Innovativeness	&&	Perceived Ease of Use&&\\\hline
\citeA{Freude2019} &	Disposition	&&&		\begin{tabular}{@{}c@{}}Institution-Based\\ Situation Normality\end{tabular}	&Knowledge-based Familiarity\\\hline
\citeA{Park2020}&&		Engagement&	Process&	Situational Normality	&\\\hline
\citeA{Wojton2020}	&&&		Understanding		&&	\\\hline
\citeA{Chi2021}&	Trusting Stance&	\begin{tabular}{@{}c@{}}Technological\\ Attachment\end{tabular}&&&			\begin{tabular}{@{}c@{}}Familiarity,\\ Social Influence\end{tabular}
\end{longtable}
\end{landscape}

\begin{landscape}
\begin{longtable}[c]{l"c|c|c|c|c}
\caption{Mapping Terms between Identified Learned Trust Factors of Significant Reliability and Validity.  Bolded authors are among those in the Top 12 survey instruments.  Entries in red were found to be reliable and valid but span more than one factor.}\\
\label{tab:termMap2}
   \textbf{Author}& \textbf{	Structural Trust} &	\textbf{	Capability-Based Trust} &	\textbf{	Affective Trust} &	\textbf{	General Trust} &	\textbf{	Intention to Use}\\\thickhline
CPRS: \citeA{IdramaniL.SinghRobertMalloy1993}&	Reliance&	Trust&	Safety&&		Confidence\\\hline
\textbf{\citeA{Madsen2000}}&&\begin{tabular}{@{}c@{}}		Perceived Reliability\\\end{tabular} &&		Faith&	\\\hline
\textbf{\citeA{Gefen2003}}&	\begin{tabular}{@{}c@{}}Institution-Based\\ Structural Assurance\end{tabular}&	Perceived Usefulness&	Calculative-Based Beliefs&	Trust&	Intended Use\\\hline
\textbf{\citeA{Korber2018}}&&		Reliability, Competence&	Intention of Developers&	Trust in Automation&	\\\hline
\citeA{Scopelliti2005}&&&&&					\\\hline
\begin{tabular}{@{}l@{}}\citeA{Wang2005}\\\citeA{Komiak2006}\end{tabular}&	Integrity&	Competence&	\begin{tabular}{@{}c@{}}Benevolence\\(Perceived Personalization)\end{tabular}	&&	\begin{tabular}{@{}c@{}}Intention to Adopt\\ Intention to Delegate\end{tabular}\\\hline
\textbf{Heerink (2009, 2011)}&&\begin{tabular}{@{}c@{}}Perceived Usefulness,\\ Attitude\end{tabular}&Perceived Adaptability &Trust&Intention to Use\\\hline
\textbf{\citeA{McKnightD2011}}&	Structural Assurance&	\begin{tabular}{@{}c@{}}Functionality,\\Reliability\end{tabular}&	Helpfulness&	\begin{tabular}{@{}c@{}}Trusting Beliefs\\ in Specific Technology\end{tabular}&	\begin{tabular}{@{}c@{}}Intention to Explore\\Deep Structure Usage\end{tabular}\\\hline
\textbf{Merritt et al. (2008, 2011a)}&&		Trust Scale	&&	Perceived Machine Accuracy&	Reliance\\\hline
\citeA{Sollner2012}	&&		Performance (formative)&	Purpose (formative)&	Trust (reflective)&	\\\hline
\textbf{\citeA{schaefer2013perception}}&	\begin{tabular}{@{}c@{}}\color{red}Robot Behaviors +\\\color{red}Communication\end{tabular}&	\begin{tabular}{@{}c@{}}Performance-Based\\ Functionality\end{tabular}	&&&	\\\hline
\citeA{Sollner2013}	&&		Predictability, Performance&	Helpfulness&	Trust in IT Artifacts&	\\\hline
Wechsung et al. (2013)&	Trust&&		Helpfulness	&&	\\\hline
Ullman et al. (2014, 2021)&	Sincere, Ethical&	Reliable, Capable&&&\\\hline			
\citeA{cherif2014impact} & 	\begin{tabular}{@{}c@{}}\color{red}Recommendation Agent\\ \color{red}Trust\end{tabular}&		\begin{tabular}{@{}c@{}}\color{red}Recommendation Agent\\ \color{red}Trust\end{tabular}&		\begin{tabular}{@{}c@{}}\color{red}Recommendation Agent\\ \color{red}Trust\end{tabular}&	Website Trust&	Intentions\\\hline
\textbf{\citeA{Chien2014}}	&\color{red}Factor 5&		\begin{tabular}{@{}c@{}}Performance Expectancy\end{tabular}&  \color{red}Factor 5& \color{red}Factor 5&	\\\hline	
\citeA{Lee2015}		&&&&& 		\\\hline		
\citeA{Tussyadiah2020}	&&	Reliability, Functionality&	Helpfulness&		Trusting& Intention\\\hline
\textbf{\citeA{Hoffman2013}}&&		\begin{tabular}{@{}c@{}}Human-Robot Fluency,\\Robot Relative\\ Contribution\end{tabular}&	\begin{tabular}{@{}c@{}}Working Alliance\\ Goal Subscale\end{tabular}&	Trust in Robot&	\\\hline
\citeA{Rupp2016}&&&			(Affective Parts of Jian)	&&	Predicted Use\\\hline
\citeA{Hegner2019}&&&&			Trust&	Adoption Intention\\\hline
\citeA{Freude2019}&	\begin{tabular}{@{}c@{}}Institution-Based\\ Structural Assurance\end{tabular}&		Calculative-Based Beliefs&&	Trust&	Intention\\\hline
\citeA{Park2020}&	Structural Assurance&	Performance&	Purpose&	Trust in Service Robots&	\begin{tabular}{@{}c@{}}Intention to Stay\\ in the Hotel\end{tabular}\\\hline
\citeA{Wojton2020}	&&	Performance	&&&		\\\hline
\citeA{Chi2021} &&	\begin{tabular}{@{}c@{}}Trustworthy Robot\\ Function and Design\end{tabular}&	\begin{tabular}{@{}c@{}}Robot-Service Fit\\Trustworthy Service Task\\ and Context\end{tabular}&	SSRIT&

\end{longtable}
\end{landscape}

\end{document}